\documentclass[a4paper]{article}

\usepackage{amsopn}
\usepackage{amssymb}
\usepackage{amsmath}
\usepackage{url}
\usepackage{graphicx}
\usepackage[nocompress]{cite}

\newcommand{\RC}[2]{\texttt{{R#1C#2}}}

\newcommand{\X}[2]{
\texttt{#1~}\hrulefill\\
\texttt{#2}
\\
\nopagebreak
\vspace{-5pt}
\hrule
\vspace{5pt}
}

\newcommand{\Y}[2]{
\texttt{*#1~}\hrulefill\\
\texttt{#2}
\\
\nopagebreak
\vspace{-5pt}
\hrule
\vspace{5pt}
}
\newcommand{\lv}{\mathrm{lv}}
\newcommand{\card}{\mathrm{card}}

\DeclareMathOperator{\semijoin}{\ensuremath{\ltimes}}
\DeclareMathOperator{\join}{\ensuremath{\bowtie}}

\begin{document}

\title{Translating Relational Queries into Spreadsheets}
\author{Jacek Sroka \and Adrian Panasiuk \and Krzysztof Stencel \and Jerzy Tyszkiewicz \\
Institute of Informatics, University of Warsaw, ul. Banacha 2,\\
02-097 Warsaw, Poland,\\
\texttt{\small\{sroka,stencel,jty\}@mimuw.edu.pl, a.m.panasiuk@riseup.net}
}

\begin{abstract}
Spreadsheets are among the most commonly used applications for data
management and analysis.  Perhaps they are even among the most widely
used computer applications of all kinds.  They combine in a natural and
intuitive way data processing with very diverse supplementary
features: statistical functions, visualization tools, pivot tables,
pivot charts, linear programming solvers, Web queries periodically
downloading data from external sources, etc. However, the spreadsheet
paradigm of computation still lacks sufficient analysis.

In this article we demonstrate that a spreadsheet can implement all
data transformations definable in SQL, without any use of macros or
built-in programming languages, merely by utilizing spreadsheet
formulas.  We provide a query compiler, which translates any given SQL
query into a worksheet of the same semantics, including NULL values.

Thereby database operations become available to the users who do not
want to migrate to a database. They can define their queries using a
high-level language and then get their execution plans in a plain
vanilla spreadsheet.  No sophisticated database system, no spreadsheet
plugins or macros are needed.

The functions available in spreadsheets impose severe limitations on
the algorithms one can implement. In this paper we offer $O(n\log^2n)$
sorting spreadsheet, but using a non-constant number of rows,
improving on the previously known $O(n^2)$ ones.

It is therefore surprising, that a spreadsheet can implement, as we
demonstrate, Depth-First-Search and Breadth-First-Search on graphs,
thereby reaching beyond queries definable in SQL-92.

\end{abstract}

\maketitle

\tableofcontents
\section{Introduction}

Spreadsheets are the desktop counterpart of databases and
OLAP in enterprise-scale computing.  They serve basically the same
purpose --- data management and analysis, but at the opposite extreme
of the data quantity scale.

Spreadsheets are very popular, and are often described as the very
first ``killer app'' for personal computers.  Today they are used to
manage home budgets, but also to create, manage and examine extremely
sophisticated models and data arising in business and research.

In his keynote talk \cite{Gates99} during SIGMOD 1998 Bill Gates
spoke about the role and challenges for spreadsheets:

\begin{quote}\it
A lot of users today find the true databases complex enough that they
simply go into either the word processor, with the table-type
capabilities, or into the spreadsheet, which I'd say is a little more
typical, and use that as their way of structuring data.

And, of course, you get a huge discontinuity because, as you want to
do database-type operations, the spreadsheet isn't set up for that.

And so then you have to learn a lot of new commands and move your data
into another location.

What we'd like to see is that even if you start out in the
spreadsheet, there's a very simple way then to bring in software that
uses that data in a richer fashion, and so you don't see a
discontinuity when you want to move up and do new things.

But that's very easy to say that. It's going to require some
breakthrough ideas to really make that possible.
\end{quote}

Despite that encouragement, relatively little research has been
devoted to spreadsheets and consequently they are still poorly
understood. In particular, 16 years later \textit{Excel} users show up
at community forums asking for help in performing database operations
on their spreadsheet data
\cite{MrExcel:Join,MrExcel:social,MrExcel:combining,MrExcel:v-lookup,MrExcel:MsQuery}.

Probably the same group of users is the target of Google.
Their spreadsheets have a very useful \texttt{QUERY} function.
It is used to run Google Visualization API Query Language
queries across data\cite{Google:QL}.
The on-line help provides the following example formula:

\texttt{=QUERY('Example Data'!\$A\$2:\$H\$7, "select B, MAX(D) group by
  B")}.  

However, this function does not permit joining relations, and
is incompatible with other spreadsheet systems.

The second notable fact is that spreadsheet language of formulas of
Excel has become a \textit{de facto} standard.  It is implemented in a
large number of spreadsheet systems, available for all major operating systems and
hardware platforms, starting from handholds and ending in the cloud,
from proprietary to open source.

Computer applications in the form of formula-only spreadsheets are
therefore highly portable, probably to the extent comparable with Java
bytecode. From this perspective, spreadsheet systems can be regarded
as virtual machines, offered by various vendors, on which spreadsheet
applications can be run.

It is therefore extremely surprising that those machines are
predominantly programmed manually, with no compilers producing
spreadsheet code from higher-level languages.

The main topic of this article is to offer a fully automated method to
construct spreadsheet implementations for a wide class of relational
data transformations. We have implemented all operators of relational
algebra, including grouping and aggregations. On top of that, we also
offer a tool to specify the transformations in a quite rich fragment
of SQL. This is our answer to the challenge posed by Bill Gates: the
discontinuity between spreadsheets and databases is reduced by the fact
that the former have the ability to express relational queries, and
the users of spreadsheets can perform relational data transformations
in the spreadsheet itself.

In the same way we address also our second point: our tool is a
compiler from a high-level language into the language of spreadsheet
formulas. The full automation of the translation process reduces the
number of human-introduced errors in the spreadsheet application, in
which the spreadsheet formulas produced in the translation are
used. As a result users can still work in the vanilla spreadsheet
environment, benefit from high portability and other features like
data analysis and visualisation, while the complex parts are generated
by a tool that allows to express them in a better suited high level
database vocabulary and avoids errors in complex computations.

\section{The contribution}

The present paper offers a twofold contribution. 

It is an extended version of an earlier paper \cite{Tyszkiewicz},
which demonstrated as a ``proof of concept'', that \textit{Excel} (and
other spreadsheets) are capable of storing and querying relational
data, and can thereby serve as database engines. In that paper
relational algebra was implemented in the spreadsheets. 

Now we extend this claim by demonstrating an automated translator,
capable of producing formulas-only spreadsheet implementations of
queries written in SQL. And indeed, we treat a spreadsheet very much
as a virtual machine, which provides a set of system functions, which
we use to implement relational queries. The resulting worksheets are
usually complex and their creation by hand could be cumbersome and
prone to errors.  Our compiler creates them without any human
intervention. When compared to the earlier paper, we add a complete
implementation of NULL values, according to the three-valued logic of
SQL.

The functionalities of spreadsheets we utilize make our
implementations work without using any plug-ins or macros. 

\textit{The reader should bear in mind, however, that our claim of
  translating SQL queries into spreadsheets does not mean, that we can
  translate the algorithms typical RDBMS systems employ to implement
  SQL. In particular, most of the algorithms we use are of quadratic
  time complexity, and hence inefficient if used on large data
  sets. Moreover, our translation tool in its present form does not
  perform optimization.}

Our investigation can also be understood as an inquiry into the
computational power of spreadsheets. In this sense, we prove that they
subsume the power of relational queries, although sometimes, as
explained, above, by algorithms inferior than those usually employed
in RDBMS. With this perspective in mind, we provide three additional,
isolated elements, absent in the earlier paper \cite{Tyszkiewicz}.

One of them is an efficient sorting algorithm, implemented by
spreadsheet formulas. The original sorting demonstrated in
\cite{Tyszkiewicz} was of quadratic time complexity. The present
algorithm is $O(n\log^2n)$. Its drawback is that it requires $4\log n$
columns to sort $n$ items. Therefore we did not decide to use it in
our automated SQL to spreadsheet translator, although 80 columns would
already suffice to sort the largest number of columns of data, which can be
stored in Excel 2013 --- the newest one at the time of this writing.
The other two algorithms we implement go beyond standard SQL. We
present a recursive implementation of Breadth-First-Search for
directed acyclic graphs, and an iterative implementation of
Depth-First-Search for arbitrary graphs. This sheds some light on the
real computational capabilities of spreadsheets, and their ability to
express recursive queries.

As the model of spreadsheet syntax and semantics we take the \textit{
  Microsoft Excel}\textsuperscript{TM} \cite{Excel}.

\begin{figure*}[htb]
\centering
\includegraphics[width=6.7in]{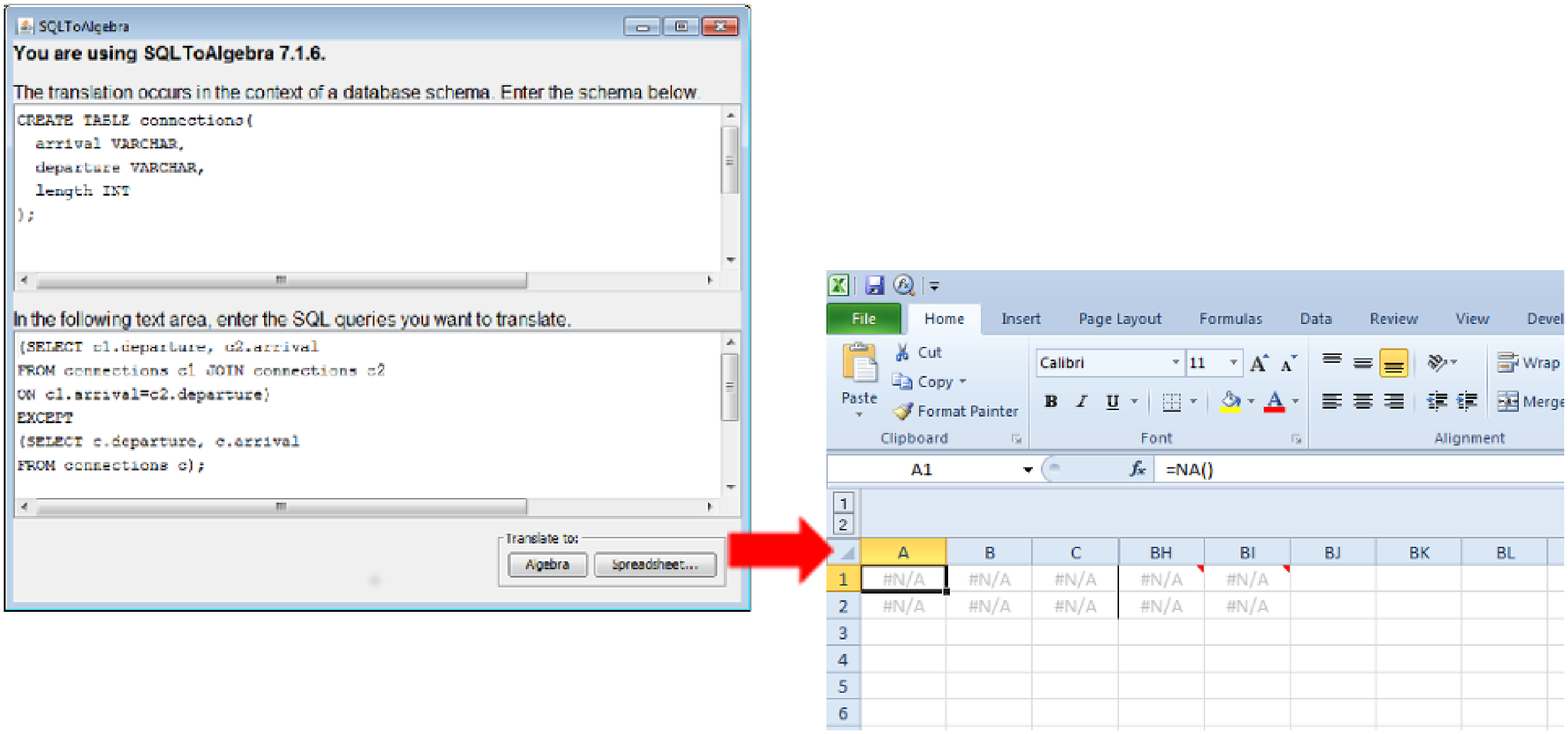}
\caption{The idea of a database implementation in a spreadsheet. A
  Java program translates an SQL query into an equivalent
  spreadsheet. Table definitions visible above are not translated, but
  determine numbers of columns of the input tables.}
\end{figure*}

\subsection{Application scenarios}

We envisage the main group of potential users of our work to be
characterized by the following:

\begin{itemize}
\item They are experienced and relatively proficient users of spreadsheets,
  mainly Microsoft Excel.
\item They are approaching the limits of spreadsheet abilities.
\item Either they are not yet ready to migrate to a database, or
  they do not want to migrate at all.
\end{itemize}

One of us (J.Ty.) was active at the \texttt{MrExcel.com} forum where
users of Excel can exchange tips and solutions, performing a kind of
\textit{participant observation}.  It has turned out that requests to
help in joining datasets are not uncommon.

One type of requests for help comes from users who are aware of
database operations and clearly state what they need, as in the threads
\cite{MrExcel:Join,MrExcel:MsQuery}.

The other type of requests comes from users, who describe the
operation they need in plain words, clearly demonstrating, that they
do not know databases.  The first discussion \cite{MrExcel:social}
concerns social research.  Its initiator needs to self-join a table on
employment in companies, to detect pairs who work together in the same
company.  This task can be very concisely formulated as a query in SQL
or the relational algebra.  The ability to compile such a query into
Excel formulas will significantly reduce the amount of necessary user
work.  In the second discussion \cite{MrExcel:v-lookup} a user needs
to join two tables of sensor data: humidity and temperature, both
keyed by date.  This meteorological application amounts to a textbook
outer join.  Matching humidity and temperature records are to be
identified and non-matched records are to be retained.  Again, the
formulation of this query in a high level language is short and
precise.  Once it gets compiled to Excel formulas, it solves the
user problem.  The topic of the third discussion
\cite{MrExcel:combining} is the problem of assembling services from
components.  The requesting user needs to join an association table of
components with the service table that relates services with
components.  This time the problem reduces to an interesting inner
join that leads to a non-4NF result.  However, it can be solved with
an inner equijoin.  Its formulation in SQL and compilation to
spreadsheet formulas solves the problem very quickly.  To summarize,
real users do need to join Excel data tables in various ways (self-,
outer-, inner-, non-NF).  A compiler of queries can be a valuable ally
in their efforts.

It is instructive to have a look at the thread \cite{MrExcel:MsQuery}.
The user wants help in performing a join, and
wants to do that in MsQuery. However, the Excel-only solution turns
out to require just two formulas per data sheet and few clicks. If
implemented with only formulas (instead of removing duplicates by a
built-in Excel menu), it would become more complex, but still perfectly
doable while the formulas would automatically recalculate without
additional clicking, if the input data would change.

\subsubsection{Spreadsheet to database migration?} 

We should explain why we think it makes sense to extend capabilities
of spreadsheets by database functionalities, what we do here, instead
of advising the users to migrate to a fully fledged database.

The reason is that spreadsheets combine in a natural and intuitive way
many very diverse features. Just to name a few, they offer pivot
tables, pivot charts, statistical functions, linear programming
solvers, Web queries periodically downloading data from external
sources, visualization tools, etc., suitable for datasets of moderate
size. Re-creating them after migration to a database would require
installing several external applications, configuring them to achieve
integration, resulting in a complex system, whose overall usability
would most likely be worse than that of the initial spreadsheet.

Finally, spreadsheets are very portable, much more than any database
system. Their unique way of combining the application code (in the
form of spreadsheet formulas) and the data in one file, and the access
to both from the common interface gives the user the ability to open,
analyse and edit them almost everywhere.

\subsubsection{Spreadsheet error rate reduction}

The second possible use of our solutions is the automated creation of
data transformation formulas for \textit{Excel} spreadsheets. They can be
inserted into existing spreadsheet applications, performing complex
data manipulations.

Today many users create such formulas manually, which results in high
error rates and incomprehensible spreadsheets. We believe, that a
typical relational query, written in SQL or relational algebra, is
significantly easier to create and understand than its spreadsheet
implementation. From this point of view, translating queries into the
language of spreadsheet formulas is translating a higher-level
language into a lower-level one.

Relational transformations can also be used to asses and improve the
quality of data in spreadsheets. Many well-known integrity constraints
can be enforced by SQL queries. The same goal can be therefore
achieved by spreadsheet translations of relational queries. If they
are inserted to an existing spreadsheet, they can indicate violations
of integrity constraints. 

\subsection{Related work}

To the best of our knowledge, the problem of expressing relational
algebra and SQL in spreadsheets has not been considered in
the setting we adopt here prior to \cite{Tyszkiewicz}.

The following results are the most similar to our work. 
The article \cite{icde}
proposes an extension of the set of spreadsheet functions by a
carefully designed database function, whereby the user can specify and
later execute SQL queries in a spreadsheet-like style, one step at a
time. These additional operators are executed by a classical database
engine running in the background. Our contribution means that exactly
the same functionality can be achieved by the spreadsheet itself. Two
papers \cite{2,3} describe a project, later named \textit{Query by Excel}
to extend SQL by spreadsheet-inspired functionality, allowing the user
to treat database tables as if they were located in a spreadsheet and
define calculations over rows and columns by formulas resembling those
found in spreadsheets. In the final paper \cite{4} a spreadsheet
interface is offered for specifying these calculations, which had to
be specified in an SQL-like code in the earlier papers. Finally,
\cite{6} describes a method to allow RDBMs to query data stored in
spreadsheets.

There is also a number of papers which discuss various methods to
support high-level design of spreadsheets, in particular
\cite{1140346,Forms,944721,1,885148,63244,1194878,184736,DBLP:conf/iccl/YoderC94}.
Some of them consider spreadsheets from the functional programming
perspective.

\subsection{Technicalities}

We assume the reader to be basically familiar with spreadsheets.  The
present article is written to make the solutions compatible with
\textit{ Microsoft Excel}, from version 2007 onward.  This version
introduced a number of new functions, absent in the earlier versions
of \textit{ Excel}. They allowed us to simplify implementations of
several operators, when compared to the conference paper
\cite{Tyszkiewicz}, which offers solutions compatible with older
versions of \textit{Excel}.

\section{Application example}

Almost every present day bank offers its customers internet access to
their accounts. One of the standard possibilities is the download of a list
of all transactions on the account in a given period. Most typically
it is an Excel file. 

So let us assume a bank account owner who wants to do a serious analysis
of their financial activities in the past year. First of all,
probably they do not make on average more than 10 bank operations
per day, which makes the data to consist of at most 4000 rows. Data of
this size can be reasonably processed in a spreadsheet.

Details of the data organization will differ from country to
country and from bank to bank, but certainly the following fields will
be provided in the file with \texttt{Transactions}:
\begin{itemize}
\item \texttt{Date} of the transaction,
\item \texttt{Trans\_type},
\item \texttt{Amount},
\item \texttt{Balance} after transaction,
\item \texttt{Card\_number} (NULL in case of non-card operations),
\item \texttt{Address} where the transaction took place.
\end{itemize}

Now the user has many options, how to process the data:
\begin{enumerate}
\item The query 
\begin{verbatim}
SELECT Address, count(Amount)
FROM Transactions
WHERE Trans_type='ATM withdrawal'
GROUP BY Address
\end{verbatim}
returns the list of addresses of ATM's used together with the
frequency of their use.
\item 
\begin{verbatim}
SELECT Card_number, sum(Amount)
FROM Transactions
WHERE Trans_type='Card payment'
GROUP BY Card_number
\end{verbatim}
returns the sums paid using each of the cards operating on the account.
\item If parents and children have accounts in the same bank, they can
  match allowances transferred by parents with their receipt by
  children,  
\begin{verbatim}
SELECT Card_number, sum(Amount)
FROM Parent_Transactions t1 JOIN
     Child_Transactions t2
ON t1.Amount=-t2.Amount AND
   t1.Date=t2.Date
\end{verbatim}
\item Self-join can be used to match a transaction creating account
  overdraft (and therefore a loan from the bank) and its
  repayments. It is however difficult to provide an SQL query here,
  since it very much depends on the details how loan identifier is
  included in the descriptions of its repayments.
\end{enumerate}
While the first two queries can be, in principle, expressed in Excel's
Pivot Table, the third and fourth query cannot be.

\section{Translation of SQL to spreadsheet}

\subsection{Architecture of a database implemented in a spreadsheet}\label{Arch}

In this article, we disregard a number of minor issues arising in a
practical implementation of the database operations in a spreadsheet.
First of all, there is the obvious limitation on the number and sizes
of relations, views and their intermediate results, imposed by the
maximal available number of worksheets, columns and rows in the
spreadsheet system at hand.  Next, the size of the data values
(integers, strings, etc.) is also limited.  The variety of data types
in spreadsheets is also restricted when compared to database systems.

The overall architecture of a relational database implemented in a
spreadsheet is as follows.  Given a specification of a query in SQL,
its implementation is created by our query compiler, in the form of an
\texttt{.xlsx} file.  The resulting spreadsheet implementation of the
query is a single worksheet, consisting of the necessary number of
columns for the data tables and, next to them, the columns performing
the computations. Initially there are always two rows of formulas, and
the user is supposed to mark the second row of the formulas and fill
with its content as many rows as necessary, taking into account the
size of the relations to be processed and the expected size of the
output. The first row should be left untouched, because sometimes it
contains formulas which differ from those which fill the remaining
rows.

The columns performing computations, and producing thereby
intermediate results, are not supposed to be edited by the user.

When the user manually enters data into the tables, the automatic
recomputation of the spreadsheet causes the output of queries to be
computed and appear in the columns with the result.

We assume the semantics over fixed domains of integers, Booleans,
texts, so that a relation is a set or multiset of tuples over these
domains, in the form implemented in the spreadsheet software.

The representation of a relation $r$ of arity $n$ is a group of $n$
consecutive columns in a worksheet, whose rows contain the tuples in
the relation. It is a crucial assumption, that the user data does not
contain spreadsheet error codes. We use those codes in our
implementation for representing special information, and they would
have been misinterpreted, if present in the initial data.

The representation of an SQL query $Q$ of arity $m$ is a group of
$m+l$ consecutive columns in a worksheet. Initial rows of those
columns are filled with formulas. In the last $m$ columns those
formulas should return either (a component of) a tuple in the result
of $Q$, or \texttt{\#N/A!} --- the \texttt{NO-DATA-NULL} (see
discussion in section \ref{NULL} below). In the remaining $l$ columns
the formulas calculate intermediate (auxiliary) results. A worksheet
of this kind can be created by entering the formulas in the first and
second row, and then \textit{filling} the second row downward to span
as many rows as necessary. This uniformity assumption means in
particular, that the formulas are completely independent of the data
they will work on.

In the following we will consider both set and bag (multiset)
semantics of the relational algebra. In the first case, duplicate rows
are not permitted in the relations and queries; in the latter they are
permitted.  However, even in the set semantics a spreadsheet
representation of a relation may contain many null rows, i.e., ones
filled with \texttt{NO-DATA-NULL} values.

Furthermore, the representation may be \textit{loose} if null rows are
interspersed with the tuples, or \textit{standard} if all the tuples come
first, followed by the null rows.

Consequently, we have loose-set, loose-bag, standard-set and
standard-bag semantics.  No matter which of the above semantics we
have in mind, the result of the query appears exactly as if it were a
table, and can be used as such. Now the only thing necessary to
compose queries is to locate their implementations side by side in a
single worksheet and change input column numbers in the formulas
computing the outermost query, to agree with the column numbers of the
outputs of the argument queries. Then the output columns of the
argument queries become the intermediate results columns of the
composition.

Therefore, queries represented in this way are compositional. We
utilize this, implementing the usual operators of relational algebra
from \cite{DB-implementation} in \textit{Excel}, and then composing
query plans of SQL queries from them. The list of implemented
operators consists of the following:

\begin{itemize}
\item Sorting,
\item Duplicate removal $\delta r$,
\item Selection $\sigma_{\theta}r$,
\item Projection $\pi_{i,j,\ldots}r$,
\item Union $r\cup s$,
\item Difference $r\setminus s$,
\item Cartesian product $r\times s$,
\item Grouping with aggregation $\gamma_{L,c}r$, where $L$ is any set
  composed from operators \texttt{SUM}, \texttt{COUNT}, \texttt{AVG},
  \texttt{MAX} and \texttt{MIN} applied to the columns of $r$, and $c$
  is a subset of the columns of $r$, over which we perform grouping,
\item Two operations specific to spreadsheets, absent in
  \cite{DB-implementation}: error trapping and standardization
  (see Section~\ref{Arch}).
\end{itemize}

Additionally, we have implemented two important operators, which can
theoretically be defined using those listed above, but deserve
independent implementations of much better performance.

\begin{itemize}
\item Semijoin $r\semijoin_\theta s$, where $\theta$ is an equality of
  two columns.
\item Join $r\join_\theta s$, where $\theta$ is an equality of
  two columns.
\end{itemize}

The present implementations of the operators make full use of
\texttt{COUNTIFS} and \texttt{SUMIFS} functions, and would not work in
the versions of the \textit{Microsoft} spreadsheet older than 2007.

We consider the query compiler accompanying the present paper, and
accessible from the Web page of the present paper (see
Section~\ref{webPage}), as the source of information about how the
operators are implemented. An additional functionality present in the
compiler is the mechanism for adapting the spreadsheet formulas to the
number of columns in the input relations.

In the earlier paper \cite{Tyszkiewicz} the operators were implemented
using \textit{array formulas}, to keep them compatible with \textit{Excel}
versions prior to 2007. The formulas were explained there in detail,
and the present ones have not changed much. 

Therefore we have decided to omit most of the descriptions of the
basic operators in this paper. A few of them, which are of more
interest, are described below in section \ref{sec:RA}.
\subsection{Query compiler}\label{SQL2XLSX}

Worksheets necessary to process queries composed of a single
relational operator are rather complex.  In case of queries composed
from multiple operators and especially multi-way joins, resulting
Excel formulas are almost intractable for humans.  Therefore, an
application of the described technology by typical clerks is hardly
possible.  In order to cater for their needs we have developed a
compiler of queries.  The compiler is implemented in Java.  The user can
define his/her query in SQL. The input consists of the query itself
and of the \texttt{CREATE TABLE} statements for the used tables. Their
purpose is to determine the number of columns of the input.

For each query, the compiler produces an empty worksheet that
implements the given query.  

The compilation is performed in two steps. 

\begin{enumerate}
\item SQL is translated into a relational algebra expression, according
  to the algorithm described in \cite{VandenBussche}.

\item This relational algebra expression is translated into a
spreadsheet, using Excel implementations of all operators of
the relational algebra, i.e. the projection, the selection, the equijoin,
the Cartesian product, grouping, basic aggregates (sum, min, max,
average, count), duplicate removal, sorting and set operations (union,
difference and intersection).
\end{enumerate}

Both steps are valid for the set semantics of SQL. This means that the
result of an SQL query and its spreadsheet implementation resulting from
the compilation, given identical tables as inputs, will produce
exactly the same sets of tuples.
 
However, the first step in the translation is not valid for the bag
semantics, if the query contains nested related subqueries. For such queries
the sets of tuples in the bag semantics and in
the spreadsheet will be the same, but their multiplicities may differ. 

The first step is valid for the bag semantics, when applied to simple
queries, which do not contain nested subqueries.  This is discussed
e.g.\ in \cite{DBLP:conf/icde/GrefenB94} where a multi-set relational
algebra has been defined. Furthermore, the authors have shown that a
number of rewrite rules that work for the classic set algebra hold
also in the bag algebra.  They have also given examples of ``set''
rules that do not migrate to the bag setting.  The article
\cite{DBLP:conf/icde/GrefenB94} proves that a translation of SQL
queries to algebra expressions is feasible, however it does not
present a ready-to-implement algorithm to do so.  Since the article
\cite{VandenBussche} offers such an algorithm (but only for the set
semantics), we have decided to use it.  This paper is in fact devoted
to the translation of SQL statements to Excel worksheets.  Therefore,
we have assumed that the peculiarities of bag and set semantics are a
minor issue.

We illustrate the operation of our compiler using an example query.
Assume a database table \texttt{connections} on railway
connections. Each of its rows describes a directed connection from the
station \texttt{departure} to the station \texttt{arrival} of the
given \texttt{length}.  We distinguish columns in algebra expressions by their
indexes.  Therefore we reference \texttt{departure} as 1,
\texttt{arrival} as 2 and \texttt{length} as 3.

\begin{figure*}[htp]
\centering
\includegraphics[scale=0.4]{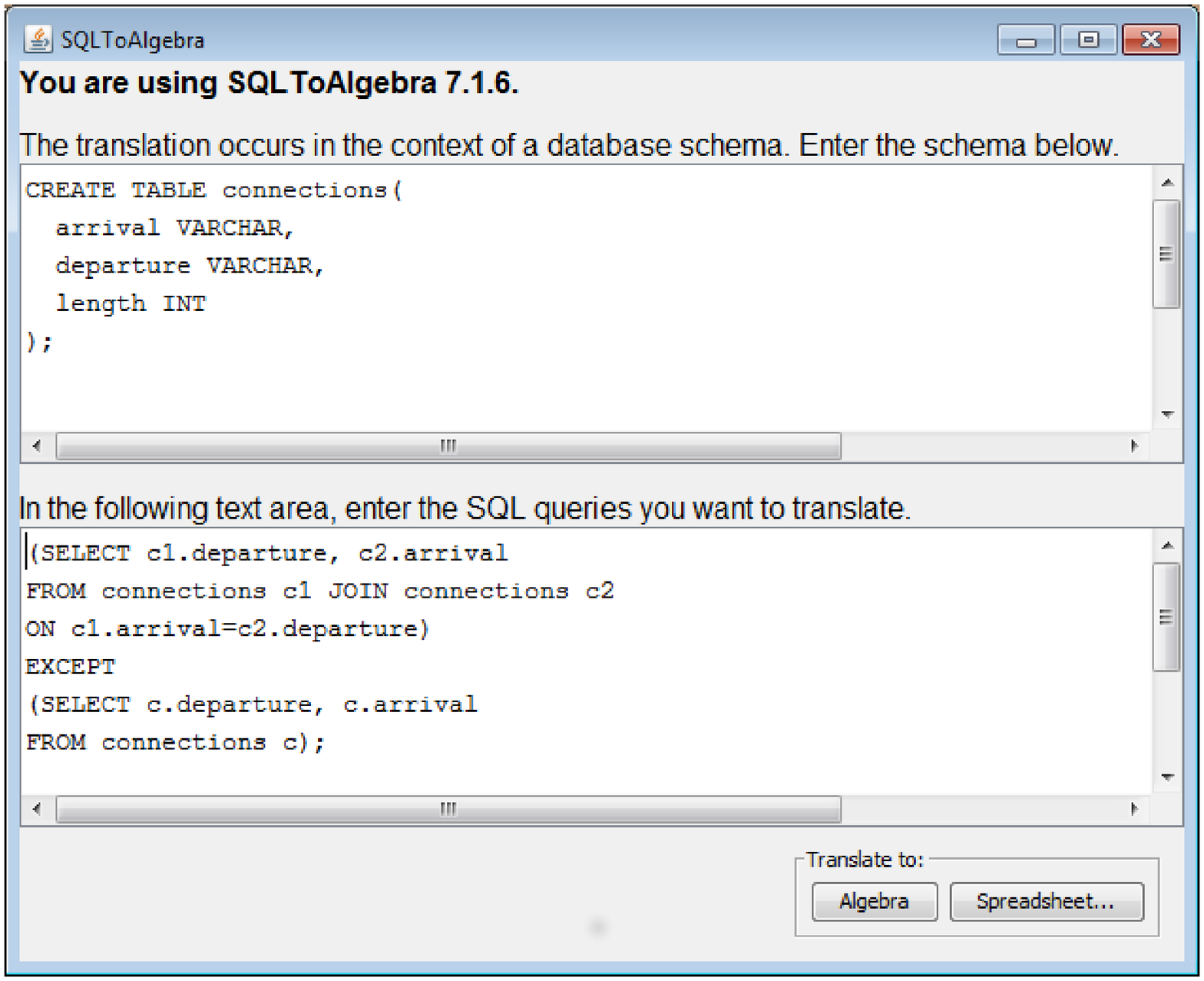}
\caption{SQL to spreadsheet compiler interface.}
\label{fig:compiler}
\end{figure*}

Let us identify pairs of stations that are not connected directly, but
the travel between them requires one change. In SQL we can formulate
this as the self join presented in Fig. \ref{fig:compiler}.

\begin{figure*}[htp]
\centering
\includegraphics[scale=0.4]{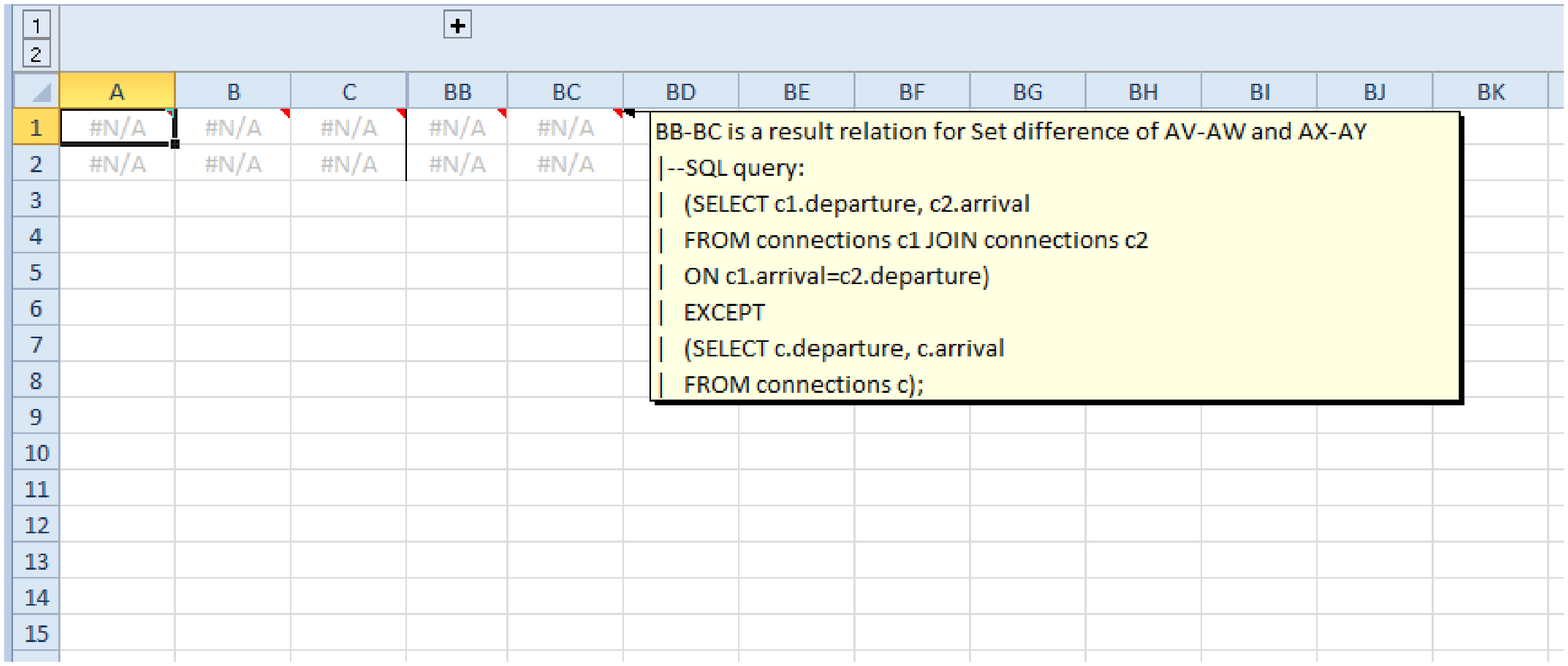}
\caption{Spreadsheet implementation of the SQL query from
  Fig.~\ref{fig:compiler}. Hovering over a cell with a small red
  triangle displays description of the column it belongs to. Columns
  between \texttt{C} and \texttt{BB} hold intermediate results and are
  hidden by default.}
\label{fig:connections}
\end{figure*}

\begin{figure*}[htp]
\centering
\includegraphics[scale=0.4]{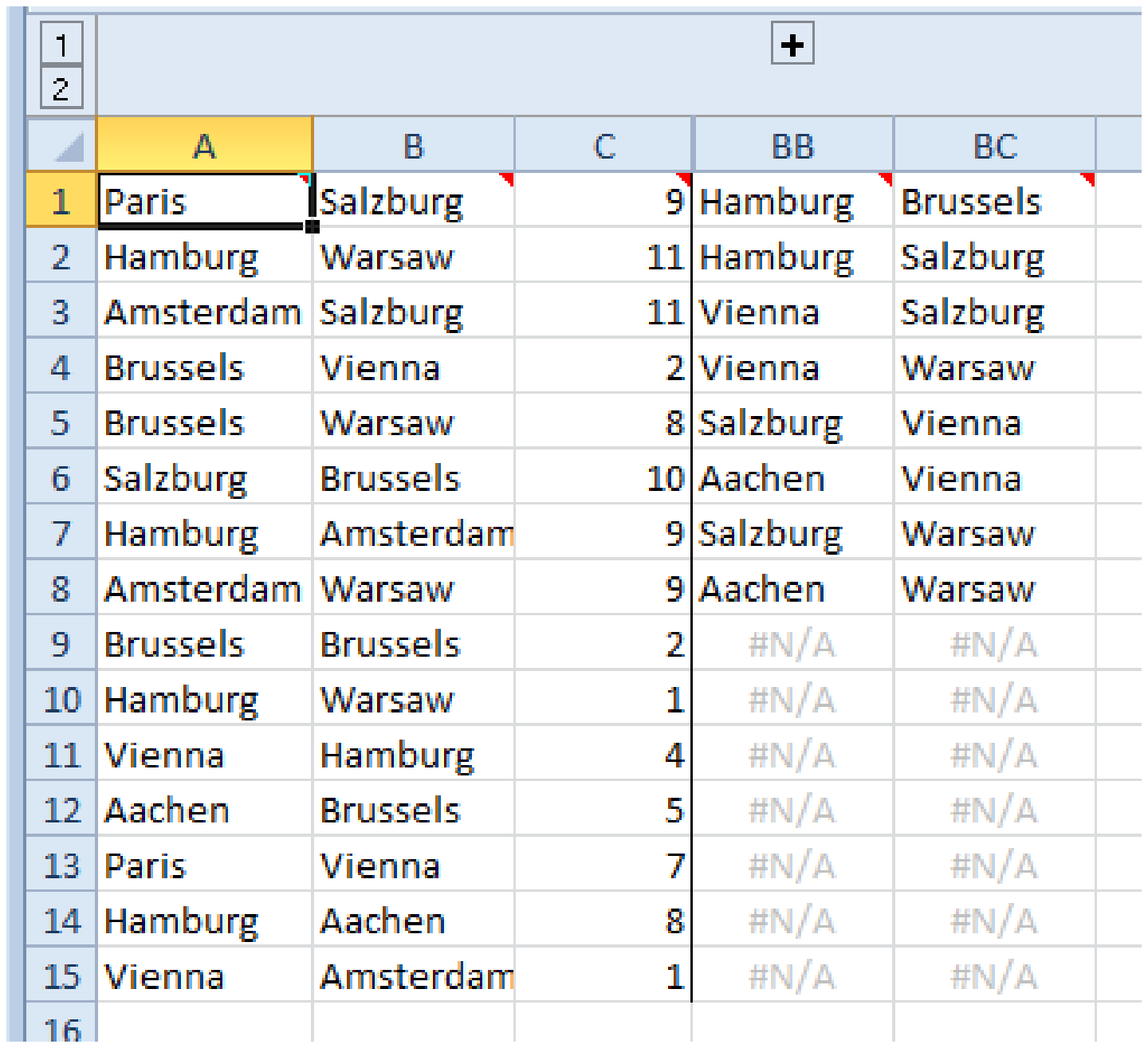}
\caption{Spreadsheet implementation of the SQL query from
  Fig.~\ref{fig:compiler} filled to 15 rows,  with sample data and
  output consisting of 3 tuples.}
\label{fig:connections2}
\end{figure*}

The first result is  translation of this query into the relational
algebra, which can be on demand shown to the user in
the following form:
\begin{verbatim}
DiffSet(
   Project(
      EqJoin(
         Reference(connections),
         Reference(connections),
         2,1
      ),
      [2, 4]
   ),
   Project(
      Reference(connections),
      [1, 2]
   )
)
\end{verbatim}

The user can also request a translation of the query into an
\textit{Excel} workbook, yielding the result shown in Figures
\ref{fig:connections} and \ref{fig:connections2}. The latter
spreadsheet results from filling the former one for the desired number
of rows and inserting the data. The user should fill sufficiently many
rows to accomodate the largest intermediate results created in the
computation process. The possibility to view the algebra expression
corresponding to the SQL query can be helpful in estimating that
number.

Cells in the first row contain comments, indicated by small red
triangles. The spreadsheet cell comments provide a very basic
explanation what the formulas compute -- indeed they identify the
relational algebra operator they implement and the columns which are
inputs to that operator.

\section{Practical Level}

This part is devoted to the discussion of the implementation issues of
our translator.
 
\subsection{NULL values}\label{NULL}

\texttt{NULL} values are unavoidable in practical implementations of
relational databases.  They lead to a three-valued logic that also has
to be implemented.  Analogously to many programming environments,
spreadsheets do not have any feature that can be easily adopted as the
database \texttt{NULL}.  Therefore we offer an implementation of
\texttt{NULL}. We will call it \texttt{NO-VALUE-NULL}.

However, we must start with \texttt{NO-DATA-NULL}, which does not
exist in standard databases, and should not be confused with
\texttt{NO-VALUE-NULL}. Indeed, the specific feature of our
implementation of database queries in Excel is that they always
process a fixed number of rows of the input tables. It does not matter
how many of them are actually occupied by data.  Hence, we must
distinguish between \texttt{NO-DATA-NULL} that means ``there is no row
of data here'' and the quite different \texttt{NO-VALUE-NULL} meaning
``there is a row of data here, but this attribute has no value''.

Our choice is to represent \texttt{NO-DATA-NULL} by the error
\texttt{\#N/A!}  which is generated by the function \texttt{NA()}.  It
has a corresponding test function \texttt{ISNA()} that returns
\texttt{TRUE} if the argument is or evaluates to \texttt{\#N/A!} and
\texttt{FALSE} otherwise.  The advantage of this representation is
that in many particular Excel formulas we use, \texttt{\#N/A!} is the
natural outcome meaning that there should be no data row at this
location.

For \texttt{NO-VALUE-NULL} we use \texttt{\#VALUE!} generated with
\texttt{=INDEX(0,-1)}.  It has the test \texttt{ISERR()}, i.e.\ a
function which returns \texttt{TRUE} if the argument is any error
\emph{except} \texttt{\#N/A!} and \texttt{FALSE} otherwise.
Unfortunately, this \texttt{NULL} does not behave exactly according to
Kleene rules when logical connectives are applied to it.  Therefore,
it requires coding of logical connectives in the queries.  However,
the test \texttt{=\#VALUE!=\#VALUE!} returns \texttt{\#VALUE!}, as
desired.

Hence, our two \texttt{NULL} representations behave as desired.  Both
of them have functions that create them and tests that distinguish
them. Those two \texttt{NULL}s are supported in the query compiler,
which we describe next.

\section{Algebra Implementation}

\subsection{\texttt{\large R1C1} notation}

In the following we use the \textit{row-column} $\RC11$-style
addressing of cells and ranges, supported by \textit{Excel}. This
notation is easier to handle in a formal description, although in
everyday practice the equivalent \texttt{A1} notation is dominating
and probably easier to understand.  Therefore our translator produces
worksheets in the \texttt{A1} notation. A user who wants to see them
in the $\RC11$ notation must change the appropriate setting in the
\textit{Excel} options.  The key advantage of the $\RC11$ notation
becomes evident when we enter a formula into a cell, click a small
handle in the lower right corner of it and extend its boundaries
either horizontally or vertically. This operation results in copying
the content of the initial cell to the new, larger area of the
worksheet. In the $\RC11$ notation the formulas resulting from filling
are identical to the initial one, which makes our explanations in the
paper much simpler.

In the \texttt{R1C1} notation, both rows and columns of worksheets are
numbered by integers starting from~1. For arbitrary nonzero integers
$i$ and $j$ and nonzero natural numbers $m,n$ the following
expressions are cell references in the \RC11\ notation: \RC{m}{n},
\RC{[i]}{m}, \RC{m}{[j]}, \RC{[i]}{[j]}, \RC{}m, \RC{}{[i]}, \RC{m}{},
\RC{[i]}{}. The number after `\texttt{R}' refers to the row number and
the number after `\texttt{C}' to the column number. If that number is
missing, it means ``same row (column)'' as the cell in which this
expression is used. A number written in square brackets is a relative
reference and the cell to which this expression points should be
determined by adding that number to the row (column) number of the
cell in which the reference is used.  A number without brackets is an
absolute reference to a cell whose row (column) number is equal to
that number.  For example, $\RC{[-1]}7$ denotes a cell which is in the
row directly above the present one in column 7, while
\RC{}{[3]}\ denotes a cell in the same row as the present one and 3
columns to the right. If \texttt{R} or \texttt{C} is itself omitted,
the expression denotes the whole column or row (respectively), e.g.,
\texttt{C7} is column number 7.

Below we discuss the most important elements of \textit{Excel} we use,
but this presentation is not exhaustive: we sometimes use functions
not presented below.

\subsection{\texttt{\large IF} function}

\texttt{IF} is a conditional function in spreadsheets.  The syntax is 

\texttt{IF(condition,true\_branch,false\_branch)}.  

Its evaluation is \textit{lazy}, i.e., after the \texttt{condition} is
evaluated and yields either \texttt{TRUE} or \texttt{FALSE}, only one of the
\texttt{branch}es is evaluated. It can be therefore used to protect
functions from being applied to arguments of wrong types, trap errors,
and, last but not least, to speed up execution of queries by avoiding
computation of certain branches.

\subsection{\texttt{\large MATCH} and \texttt{\large INDEX} functions}\label{match}

We mostly use \texttt{MATCH} using the syntax \texttt{MATCH(cell,range,0)}. It
returns the relative position of the first value in \texttt{range} which
is equal to the value in \texttt{cell}, and an \texttt{\#N/A!} error if
such a value does not appear there.

A call \texttt{MATCH(range,cell,1)} is correct only if \texttt{range} is
sorted in ascending order.  Then, such a call to \texttt{MATCH} returns
the relative position of the largest value in \texttt{range} that is less
than or equal to the value in \texttt{cell}.

\texttt{INDEX} is used in the syntax \texttt{INDEX(range,cell)}.
This function call returns the value from \texttt{range} whose relative
position is given by the value from \texttt{cell}.

\subsection{\texttt{\large COUNTIFS}}

In \textit{Excel} 2007 two new highly expressive functions \texttt{COUNTIFS}
and \texttt{SUMIFS} appeared for the first time.  The function \texttt{COUNTIFS} counts rows (columns, resp.) satisfying multiple criteria,
which can refer to several columns (rows, resp.).  The general syntax
is

\texttt{COUNTIFS(rng$_1$,cr$_1$,\ldots,[rng$_k$,cr$_k$])},

where each of the ranges \texttt{rng} has the same dimension.

If the input ranges are columns, the function returns the number of
rows $r$ such that the $r$th value in \texttt{column$_i$} satisfies
criterion \texttt{cr$_i$} for $i=1,\ldots,k$.  In the dual form, the
calculation proceeds in the same way, except that rows take over the
role of columns and \textit{vice versa}.

\begin{figure}
\centering
\includegraphics[scale=0.58]{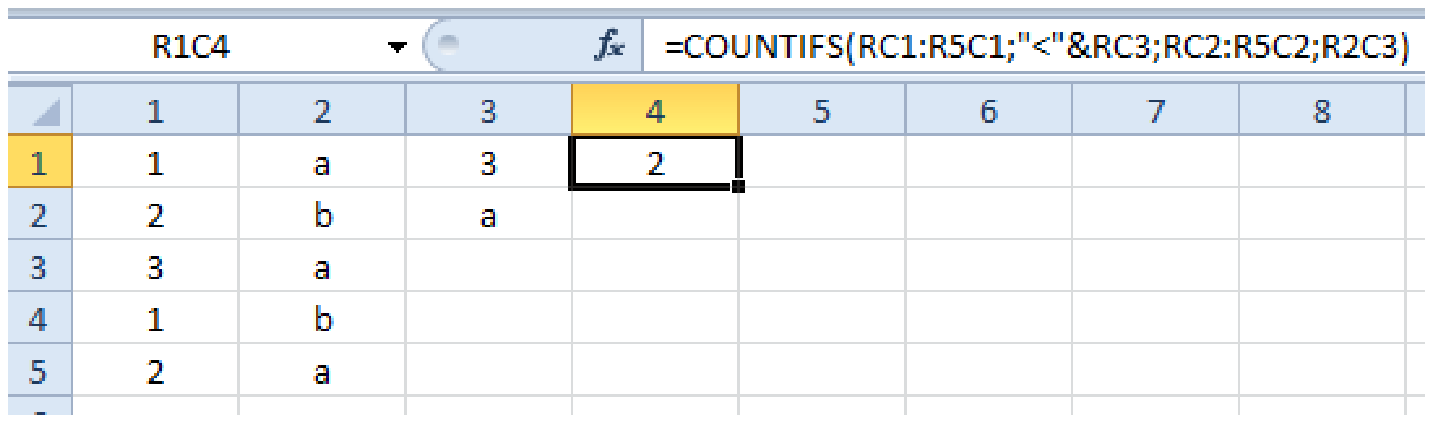}
\caption{In the context of this worksheet the formula
\protect\\ 
\texttt{=COUNTIFS(RC1:R5C1,"<"\&RC3,RC2:R5C2,R2C3)} returns 2.}
\label{countifs}
\end{figure}

For example, in the context of the worksheet depicted in Fig.
\ref{countifs}, the formula

\texttt{=COUNTIFS(RC1:R5C1,"<"\&RC3,RC2:R5C2,R2C3)}

returns 2, since there are two rows in the data, in which a number
smaller than 3 is accompanied by the string \texttt{"a"}. Note the way
of creating criteria, which can be expressed by a string produced by
concatenating (operator \texttt{\&}) the inequality sign with the
reference to the numerical argument, or by a value or reference, in which
case the condition is, by default, equality.

\subsection{\texttt{\large SUMIFS}}

This time the syntax is 

\texttt{SUMIFS(sum\_rng,rng$_1$,cr$_1$,\ldots,[rng$_k$,cr$_k$])}.

The operation of \texttt{SUMIFS} is quite similar to that of
\texttt{COUNTIFS}, except that the rows are not counted, but the
values in column \texttt{sum\_rng} are summed over the rows satisfying
the criteria. 

\subsection{\texttt{\large ROW} and \texttt{\large COLUMN}}

The function call \texttt{ROW()} returns the row number in which
the call is located. Similarly, \texttt{COLUMN()} returns the column
number in which the call is located.

\subsection{\texttt{\large OFFSET}}

\texttt{OFFSET} is a function which differs significantly from all
other functions mentioned in this paper. It allows the user to specify
an arbitrary range of cells, whose location is determined by the
numerical arguments of \texttt{OFFSET}. 
The syntax is 

\texttt{OFFSET(reference, rows, cols, [height], [width])}

The call to this function yields a range, which is determined as
follows: its top left corner is located \texttt{rows} down and
\texttt{cols} to the right of \texttt{reference}. If no optional
argument is provided, the range is a single cell, if they are present,
they specify the dimensions of the range. The specific feature of
\texttt{OFFSET} calls is that their arguments are calculated at
runtime and only then it becomes known, what cell each \texttt{OFFSET}
call refers to. In particular, when this function is used, it is
impossible to determine, before executing the spreadsheet, if circular
references are created or not.

\subsection{Notation}

We use the following convention for presenting our implementations:

\X{COLUMNS}{\texttt{=FORMULA}} 

means that \texttt{=FORMULA} is entered into
each cell of the \texttt{COLUMNS}, which may be specified either to be a
single column (e.g. \texttt{C5} ) or a range of a few columns (e.g. \texttt{
  C5:C7}), or a single cell (e.g. \texttt{R1C5}), and in each case
belongs to the columns with intermediate values. In few cases we
first fill a complete column with formulas, and then override the
formula in the first row. This is used to create columns in which the
first formula differs from those in the subsequent rows.  

The notation

\Y{COLUMNS}{\texttt{=FORMULA}} 
indicates that formulas located in \texttt{
  COLUMNS} calculate the output of the query.

\subsection{Relational algebra}\label{sec:RA}

In the examples presented here the arguments of the algebra operators are
binary or ternary relations. Our query compiler uses parameterized
forms of the implementations below, which work for any numbers of
column in the input.

Except sorting, in all other cases we assume the input to be in the
standard form, i.e., null rows are at the bottom.

We describe the following implementations, from the full list of
\cite{DB-implementation}:

\begin{itemize}
\item  Sorting;
\item Grouping with aggregation $\gamma_{L,c}r$, where $L$ is any
  combination of operators \texttt{SUM}, \texttt{COUNT}, \texttt{AVG},
  \texttt{MAX} and \texttt{MIN} applied to the columns of $r$ and $c$
  is a subset of the columns of $r$;
\item Semijoin $r\semijoin_\theta s$, where $\theta$ is an equality of
  two columns;
\item Join $r\join_\theta s$, where $\theta$ is an equality of
  two columns.
\end{itemize}

\subsection{Sorting\label{sec:sorting}}

Now we describe an implementation of sorting, used in our
translator. It is of quadratic complexity. We describe a significantly
faster sorting algorithm in section \ref{Sort} below. However, it
uses a non-constant number of columns, hence was not incorporated
into the translator.

We assume that columns \texttt{C1:C3} contain the source data and we
sort in ascending order by the values in \texttt{C1}.

\X{R1C4}{=COUNTA(C1)-COUNTIFS(C1,NA())}

\X{C5}{=IF(ISNA(RC1),R1C4+1,IF(ISERR(RC1),\\
1+R1C4-COUNTIFS(C1,RC1),\\
COUNTIFS(C1,"<"\&RC1)+\\
COUNTIFS(R1C1:RC1,RC1)))}

\X{C6}{=MATCH(ROW(),C5,0)} 

\Y{C7:C9}{=INDEX(C[-6],RC6)}

For descending sort the only change is

\X{C5}{=IF(ISNA(RC1),R1C4+1,IF(ISERR(RC1),\\
1+R1C4-COUNTIFS(C1,RC1),
\\COUNTIFS(C1,">"\&RC1)+
\\COUNTIFS(R1C1:RC1,RC1)))}  

In either case, the formulas compute in \texttt{RC5} the number of
entries in column \texttt{C1} which are smaller than or equal to
\texttt{RC1} plus the number of entries equal to \texttt{RC1} in
\texttt{R1C1:RC1}.  This is the number of the row into which
\texttt{RC1} should be relocated during sort.

Now \texttt{RC6} contains the number of the row in which the number of
the present row appears in \texttt{C5}.

Finally, the formulas in \texttt{RC7, RC8, RC9} fetch the values from
columns \texttt{C1, C2, C3} from the row calculated in \texttt{RC6}.

An important property of this operation is that this form of sorting
is stable.

\subsection{Grouping with aggregation}

Grouping is an operator, whose implementation in the translator
differs significantly from the one given in \cite{Tyszkiewicz}. The
reason is that the present one computes simultaneously many aggregates
for one grouping, unlike the former one.

In the following, we assume that the relation to be processed is
located in \texttt{C1:C10}. We wish to express the grouping, which in
SQL can be declared as follows:
\begin{verbatim}
SELECT C1,C2,MIN(C3), MAX(C4), SUM(C5),
       COUNT(C6,C7), AVG(C8),
       COUNT(DISTINCT C9,C10)
FROM C1:C10
GROUP BY C1,C2
\end{verbatim}

We have the following problems: 
\begin{itemize}
\item For each separate grouping performed we leave one
row from each group:
\begin{itemize}
\item for \texttt{MIN(C3)} and \texttt{MAX(C4)} we leave the row where
  the actual minimum or maximum is attained;
\item for the remaining operators we leave the very first row of each
  group together with the computed aggregate.
\end{itemize}
\item Now we may have between 1 and 3 entries for each group, with
  different aggregations, which must be unified to produce a single
  row with all aggregates.
\end{itemize}

\X{C11}{=COUNTIFS(R1C1:RC1,RC1,R1C2:RC2,RC2)}

This counts which occurrence of the values in \texttt{C1:C2} we have
in the present row. 

\X{C12}{=IF(ISNA(RC1),NA(),IF(ISERR(RC3),\\
COUNTIFS(C1,RC1,C2,RC2)-\\
COUNTIFS(C1,RC1,C2,RC2,C3,RC3),\\
COUNTIFS(C1,RC1,C2,RC2,C3,"<"\&RC3)))}

This is a help formula for minimum. It counts how many tuples in
\texttt{C1:C3} have the same values in \texttt{C1:C2} as in the
present row, and a smaller value in \texttt{C3}. There is a special
treatment if there is \texttt{NO-VALUE-NULL} in \texttt{C3}.

\X{C13}{=IF(ISNA(RC1),NA(),IF(ISERR(RC4),\\
COUNTIFS(C1,RC1,C2,RC2)-\\
COUNTIFS(C1,RC1,C2,RC2,C4,RC4),
\\COUNTIFS(C1,RC1,C2,RC2,C4,">"\&RC4)))}

This is an analogous formula for maximum.

\X{C14}{=IFERROR(RC[-9],"")}

\X{C15}{=IFERROR(RC[-7],"")}

These two formulas replace NULLs of both kinds by empty texts for
\texttt{SUM} and \texttt{AVG} aggregations. The reason is that
\textit{Excel}'s \texttt{SUMIFS} function produces an error when one
of its arguments is an error, but fortunately ignores text arguments.

\X{C16}{=IF(AND(ISERROR(RC[-7]),\\
ISERROR(RC[-6])),INDEX(0,-1),\\
COUNTIFS(R1C1:RC1,RC1,R1C2:RC2,RC2,\\
R1C9:RC9,RC9,R1C10:RC10,RC10))}

This help formula for \texttt{COUNT DISTINCT} counts which occurrence
of the values in \texttt{C1:C2,C9:C10} we have in the present
row. Note that it produces \texttt{NO-VALUE-NULL} in case there are NULLs in
columns \texttt{C9:C10} of the present row.

\Y{C17:C18}{=IF(RC11=1,RC[-16],NA())}

This formula turns non-first occurrences of pairs from \texttt{C1:C2}
into \texttt{NO-DATA-NULL}. All subsequent formulas test if this value in
\texttt{NO-DATA-NULL} and if so, become \texttt{NO-DATA-NULL}, too.

\Y{C19:C20}{=IF(ISNA(RC17),NA(),\\
SUMIFS(C[-16],C1,RC1,C2,RC2,C[-7],0))}

This formula relocates the minimum and maximum values (recognized by 0 in
columns \texttt{C12} and \texttt{C13}, resp.) into the present row.

\Y{C21}{=IF(ISNA(RC17),NA(),\\
SUMIFS(C[-7],C1,RC1,C2,RC2))}

This formula computes the \texttt{SUM} aggregation.

\Y{C22}{=IF(ISNA(RC17),NA(),\\
COUNTIFS(C1,RC1,C2,RC2)-\\
COUNTIFS(C1,RC1,C2,RC2,
\\C[-16],INDEX(0,-1),\\
C[-15],INDEX(0,-1)))}

This formula computes the \texttt{COUNT} aggregation, where
double \texttt{NO-VALUE-NULL} values are not taken into account. 

\Y{C23}{=IF(ISNA(RC17),NA(),
\\SUMIFS(C[-8],C1,RC1,C2,RC2)/\\
COUNTIFS(C1,RC1,C2,RC2))}

This formula computes the \texttt{SUM} aggregation.

\Y{C24}{=IF(ISNA(RC[-7]),NA(),\\
COUNTIFS(C1,RC1,C2,RC2,C[-8],1))}

This final formula computes the \texttt{COUNT DISTINCT} aggregate. The
first occurrences are indicated by 1 in column \texttt{C16}.

\subsection{Semijoin}

Assume that we are given two relations located in \texttt{C1:C2} and
\texttt{C3:C4}, respectively, and we wish to compute the equisemijoin
$\mathtt{C1:C2} \semijoin_{\mathtt{C1}=\mathtt{C3}}\mathtt{C2:C3}.$

This is achieved in the following way.  The two formulas below copy
the rows from \texttt{C1:C2}, replacing those which do not belong to
the semijoin by \texttt{NO\_DATA\_NULL}s. The resulting relation is
loose and should be normalized.

\X{C5}{=IF(ISERROR(MATCH(RC1,C4,0)),NA(),RC1)}

\X{C6}{=IF(ISNA(RC5),NA(),RC2)}

\subsection{Join}

Let two relations be located in \texttt{C1:C2} and \texttt{C3:C4},
respectively, and the equijoin
$\mathtt{C1:C2}\join_{\mathtt{C1}=\mathtt{C3}}\mathtt{C3:C4}$ should
be computed.

It is computed according to the decomposition 
\begin{eqnarray*}
&\mathtt{C1:C2}\join_{\mathtt{C1}=\mathtt{C3}}\mathtt{C3:C4}=\\
&(\mathtt{C1:C2}\semijoin_{\mathtt{C1}=\mathtt{C3}}\mathtt{C3:C4})
\join_{\mathtt{C1}=\mathtt{C3}}
(\mathtt{C3:C4}\semijoin_{\mathtt{C3}=\mathtt{C1}}\mathtt{C1:C2}) 
\end{eqnarray*}
Let us note that we have
 \[\pi_\mathtt{C1}(\mathtt{C1:C2}\semijoin_{\mathtt{C1}=\mathtt{C3}}\mathtt{C3:C4})=
\pi_\mathtt{C3}(\mathtt{C3:C4}\semijoin_{\mathtt{C3}=\mathtt{C1}}\mathtt{C1:C2}).\]
Then, denoting this the set of elements in this common projection by $
X,$ we can further decompose the join into the sum

\[
\bigcup_{x\in X }\{x\}\times
\pi_{\mathtt{C2}}(\sigma_{\mathtt{C1}=x}(\mathtt{C1:C2}))\times
\pi_{\mathtt{C4}}(\sigma_{\mathtt{C3}=x}(\mathtt{C3:C4})).
\]

We will call the expressions
$\{x\}\times\pi_{\mathtt{C2}}(\sigma_{\mathtt{C1}=x}(\mathtt{C1:C2}))\times
\pi_{\mathtt{C4}}(\sigma_{\mathtt{C3}=x}(\mathtt{C3:C4}))$ \textit{blocks}
of the join.

The \textit{Excel} implementation of the above idea is as follows.

\begin{figure}
\centering
\includegraphics[scale=0.28]{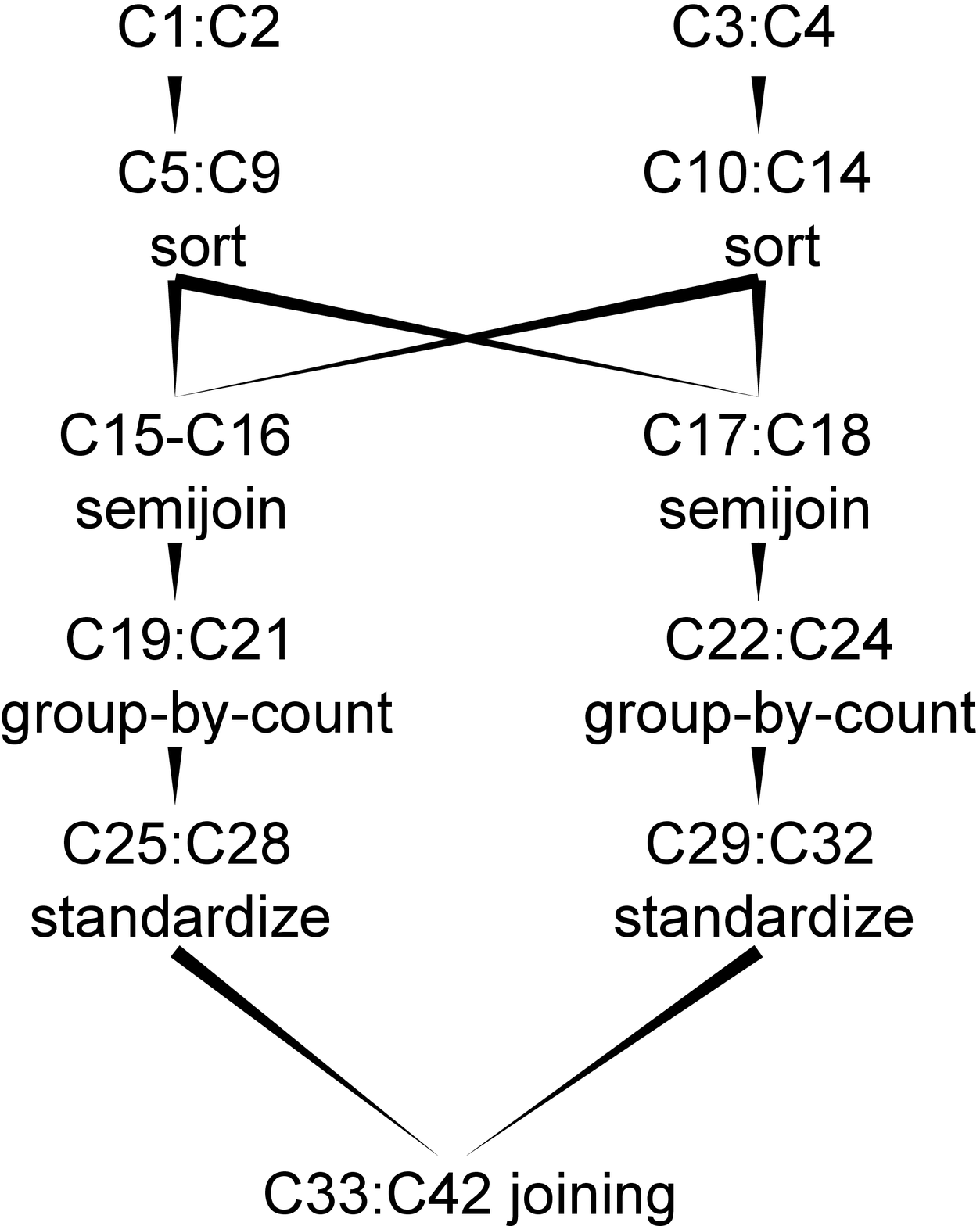}
\caption{The structure of equijoin of two relations stored in
  \texttt{C1:C2} and \texttt{C3:C4}}\label{join}
\end{figure}

The initial phase of computing the join consists of application of
several other operators, according to Fig.~\ref{join}.

After this initial phase, columns \texttt{C27:C28} contain
\[\bigcup_{x\in X }\{\langle x,\card(\sigma_{\mathtt{C1}=x}(\mathtt{C1:C2}))\rangle\}\]
and columns \texttt{C31:C32} contain
\[\bigcup_{x\in X }\{\langle x,\card(\sigma_{\mathtt{C3}=x}(\mathtt{C3:C4}))\rangle\}.\]
The real join creation takes place in columns \texttt{C33:C42} and is
executed as follows. 

\X{C33}{=MATCH(RC[-6],C[-25],0)}

\X{C34}{=MATCH(RC[-3],C[-21],0)}

These two columns contain pointers to the first rows of the sorted
input tables with consecutive elements of $X.$

\X{C35}{=IFERROR(RC[-7]*RC[-3],"")}

The values in \texttt{C35} are the sizes of the blocks of the join.

\X{C36}{=IFERROR(R[-1]C[-1]+R[-1]C,"")}

\X{R1C36}{=0}

{C36} contains the numbers of rows at which the consecutive blocks of the join
should begin minus 1. 

\X{R1C37}{=SUM(C[-2])}

This number is the total cardinality of the join to be produced.

\X{C38}{=IF(ROW()>R1C37,NA(),\\
MATCH(ROW()-1,C36,1))}

In this line function \texttt{MATCH} is used with the last parameter
\texttt{1} to do inexact search for the number of the block from which
the present tuple in join should originate.

\X{C39}{=IF(ISNA(RC[-1]),NA(),\\
IF(RC[-1]<>R[-1]C[-1],1,1+R[-1]C))}

\X{R1C39}{=IF(ISNA(RC[-1]),NA(),1)}

These two lines compute, within each block, the number of the present
row of the join within its block.

\Y{C40}{=INDEX(C[-13],RC[-2])}

\Y{C41}{=INDEX(C[-32],INDEX(C[-8],RC[-3])+\\
MOD(RC[-2]-1,INDEX(C28,RC[-3])))} 

\Y{C42}{=INDEX(C14,INDEX(C34,RC[-4])+\\
QUOTIENT((RC[-3]-1),INDEX(C28,RC[-4])))} 

In the last three columns we fetch the right $x\in X,$ then the
relevant element of $\sigma_{\mathtt{C1}=x}(\mathtt{C1:C2})$ and the
relevant element of $\sigma_{\mathtt{C3}=x}(\mathtt{C3:C4}),$ creating
that tuple of the join.

\section{Special algorithms}

The algorithms whose implementations we describe in this section are
not used by the query compiler. They arose from our attempts to
either find better algorithms for SQL-92, or to express computations
which go beyond SQL-92. 

\subsection{Fast sorting}\label{Sort}

The solution to the problem of sorting presented in Section 
\ref{sec:sorting} is clearly a quadratic algorithm.

In this section we describe an efficient linearithmic implementation.
We adopt the well-known bottom-up merge-sort algorithm and simulate a
sorting network over the data range.  

A spreadsheet with an implementation of our algorithm is available at
the Web page of the paper (see Section~\ref{webPage}). It includes two
worksheets one with a condensed version with fewer columns per step
and one with many auxiliary columns for easy human
understandability. Below we refer to the latter spreadsheet.

In the first step, pairs of neighbouring cells are sorted. The next
step leaves sorted quadruples, and so on, until the whole data range
is sorted. Obviously, if $n$ items are to be sorted, $\log n$ steps
are needed. In the spreadsheet model of computation we cannot
reiterate values in a cell. Thus our fast sorting does not work in
place, but rather uses a logarithmic number of columns,
i.e. successive columns are needed for successive steps. Such an
organization can be compared to a sorting network.

\begin{figure*}[htb]
\centering
\includegraphics[scale=0.75]{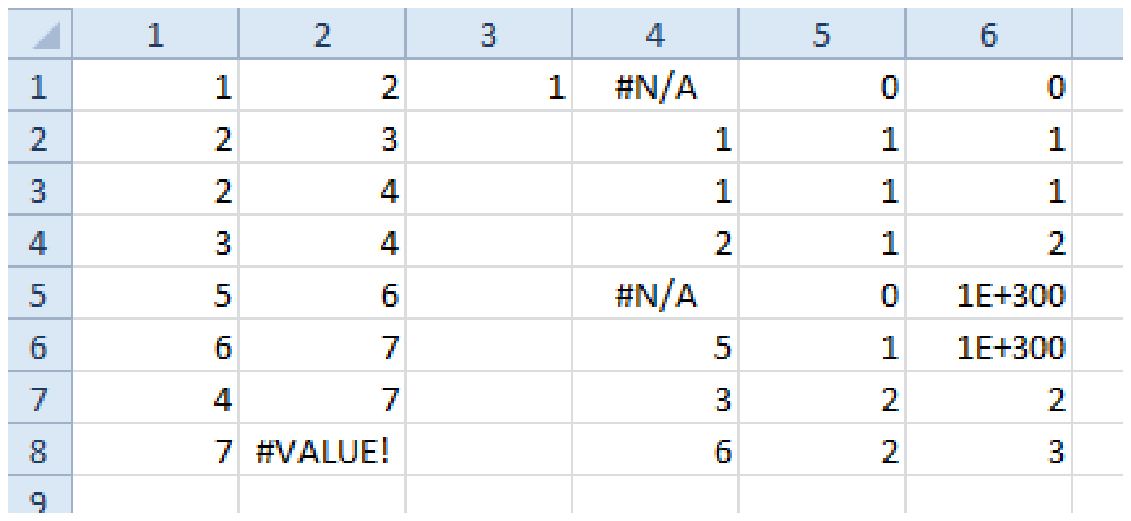}
\caption{Spreadsheet implementation of the bottom-up merge sort
  algorithm in \textit{Excel}. The area shown merges two already
  sorted 4-element blocks on the left to yield a sorted 8-element
  block on the right.}
\label{fig:merge_sort}
\end{figure*}

The fast sorting algorithm presented below is composed of 12 columns
per one level of merging. In practical situation one may condense this
to just 4 columns, replacing references to cells by formulas which are
in those cells (this process we call inlining).

Now we describe one such block. Let us assume that the already sorted
blocks of certain length are located in column \texttt{C25}. We want
to produce sorted blocks of doubled size in column \texttt{C37}. 

In order to sort $n$ items it is necessary to use $\log n$ such
groups of columns. The already sorted blocks of the initial data are
of size 1, and each group of columns merges pairs of already existing
blocks to produce sorted blocks of duplicate size.

Now we start the description of such a group of 12 columns.

\X{C26}{=QUOTIENT(COLUMN(),12)}
This column determines which level of merging is performed now. 

\X{C27}{=POWER(2,RC[-1])}
This column determines the sizes of blocks to be merged. 

\X{C28}{=QUOTIENT(ROW()-2,RC[-1]*2)*2*RC[-1]+1}
Now we determine where the top of the two blocks to be merged
starts\ldots 

\X{C29}{=RC[-1]+RC[-2]-1}
\ldots and where it ends\ldots 

\X{C30}{=RC[-2]+RC[-3]}
\ldots and where the bottom block starts\ldots

\X{C31}{=RC[-2]+RC[-4]}
\ldots and where it ends. 

The following 5 columns do the merging, and now the pattern of
references becomes more complicated. Formulas do not refer only to
cells to the left, but also to values above them, in those 5 columns.

\X{C32}{=IF(MOD(ROW()-2,RC[-5]*2)=0,\\RC[-4],IF(R[-1]C[4],R[-1]C+1,R[-1]C))}
This column computes the position of the first not-yet-merged element
from the top block. It either repeats the value \texttt{R[-1]C}
directly above in the same column, or increments it by 1, depending on
the value \texttt{R[-1]C[4]} one row above and 4 columns to the right,
which determined if the value in (then) subsequent step should be
taken from the top or bottom block. The initial test
\texttt{IF(MOD(ROW()-2,RC[-5]*2)=0\ldots} checks if this row starts a
new merging of two new blocks; if it is so then the formula returns
the top element of the new top block to be merged \texttt{RC[-4]}.

\X{C33}{=IF(MOD(ROW()-2,RC[-6]*2)=0,\\RC[-3],IF(R[-1]C[3],R[-1]C,R[-1]C+1))}
Now we do the same for the bottom block.

The following two columns retrieve the data elements from the just
determined positions.  

\X{C34}{=INDEX(C[-9],RC[-2]+1)}

\X{C35}{=INDEX(C[-10],RC[-2]+1)}

\X{C36}{=IF(RC[-4]>RC[-7],FALSE,\\IF(RC[-3]>RC[-5],TRUE,RC[-2]<=RC[-1]))}
In this column the comparison \texttt{RC[-2]<=RC[-1]} determines which
of the found values is smaller and should go now to the output, and
the result is represented as a Boolean value, to be used in the
following row of columns \texttt{C32} and \texttt{C33}. The initial
two tests verify if all values from one of the merged blocks
have already been used and we should take a value from the other
block, irrespectively of anything else.

\X{C37}{=IF(RC[-1],RC[-3],RC[-2])}
And here the chosen element, smaller of the two, is appended to the
result. 

Concerning the complexity of this implementation, for sorting $n$
items it consists of $n$ rows of formulas, and logarithmic number of
columns, $O(n\log n)$ formulas in total. The formulas in turn either
access a constant number of neighbouring cells and, in some of the
columns, call \texttt{INDEX} function to return a particular element
from the input column of unsorted items. Theoretically, the complexity
of this operation is $O(\log n),$ resulting in $O(n\log^2 n)$ total
complexity. However, tests we have performed indicate that the time
necessary to execute \texttt{INDEX} in \textit{Excel} does not depend
on the number of data elements in the input range, so in practice our
sorting works in $O(n\log n)$ time, for $n$ not exceeding the number
of rows available in \textit{Excel}.

\subsection{Graph traversing}\label{TC}

Graph traversing is a fundamental algorithmic operation. It appears as
an important step in numerous graph algorithms. Its two most important
variants are Breadth-First-Search (BFS) and Depth-First-Search (DFS).

We are going to demonstrate that both of them can be implemented in
spreadsheets, with the restriction that BFS works for acyclic
directed graphs, while DFS for arbitrary directed graphs.
An implementation of BFS for cyclic graphs can be done similarly to DFS.
However, we have decided to show the BFS version for acyclic graphs to simplify this algorithm and its presentation.
Furthermore, as we will show, this gives us an unusual way to test if a directed graph is cyclic or not.
This test is another evidence that the potential of the spreadsheet systems has not been well recognized yet. 

Technically speaking, we will demonstrate, how to \textit{order}
vertices of a graph, given as a list of edges, exactly in the order in
which BFS and DFS visit them.

In both cases, we assume that the input (directed) graph $G$ is
specified by its set $E$ of directed edges, which are pairs of
vertices. The vertex set $V$ is determined by the set of edges. We are
also given a start vertex $s\in V$, from which the traversal begins.

\subsection{Breadth-First-Search}

The idea of the algorithm is that we assign a level $\lv(v)$ to each
vertex $v\in V$.
\begin{itemize}
\item $\lv(s)=0$ for the source vertex $s$.
\item For every vertex $v\neq s$ with no incoming edges $\lv(v)=\infty$. 
\item For every other vertex $v$, we define
  $\lv(v)=1+\min\{\lv(w)~|~\langle w,v\rangle\in E\}$
  (assuming that $1+\infty=\infty$).
  \end{itemize}

Every linear ordering of those vertices of $G$,
which have finite value of $\lv$ and such that $\lv$ is
non-decreasing, corresponds to some BFS traversal of the graph.

Sorting is doable in spreadsheets, as well as filtering, hence it is
sufficient to show how to compute $\lv$.

We have one technical problem: $\infty$ is not present in
spreadsheet arithmetic. Instead we use $10^{300}$, which is more than
the number of particles in the visible part of the universe, and, by
the definition of the level, the finite values of level cannot
achieve this quantity. So it is a fully functional substitute of
$\infty$ for our purpose. 

\begin{figure}
\centering
\includegraphics[scale=0.7]{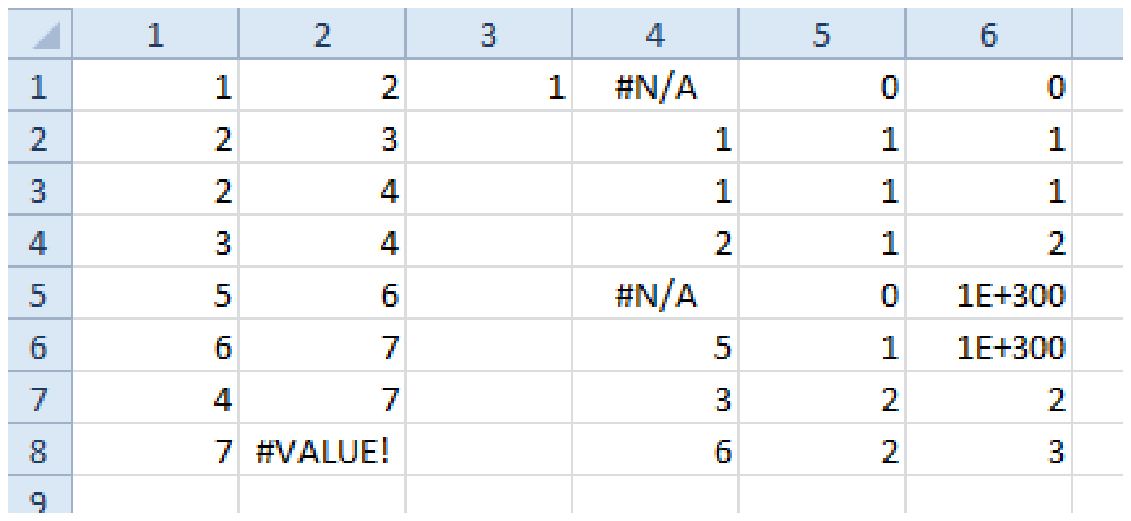}
\caption{Spreadsheet implementation of the BFS graph traversal.}
\label{fig:BFS}
\end{figure}

Now we discuss the implementation of our idea. We start with producing
an expanded set of edges by adding to $E$ edges $\langle
v,\mathtt{NULL}\rangle$ for all $v\in V$ with no outgoing edges. This
modification does not alter the values of level resulting from the
algorithm, and simplifies our task, since now every vertex appears as
a first coordinate of some edge. This operation is expressible in SQL,
hence it can be implemented in a spreadsheet, and we omit the formulas
which perform it.

Next, we sort the expanded set of edges by the second
coordinate. As a consequence the edges incoming into a vertex $v$
always form a contiguous block. Again we omit the
formulas which perform the sorting (see Section \ref{sec:sorting}).

Let the sorted edges of the expanded $E$ be provided row-wise in
columns \texttt{C1:C2} and the start node $s$ be located in cell
\texttt{R1C3}.

Each row is therefore associated with an edge $\langle v,w\rangle$
located in columns \texttt{C1:C2}, and we are going to compute
$\lv(v)$ in column \texttt{C6}.

We start by computing for each row corresponding to edge $\langle v,w\rangle$,
the (position of the) beginning of the block of all edges incoming into $v$. 
If $v$ has no incoming edges, the result is \texttt{\#N/A!}.

\X{C4}{=MATCH(RC1,C3,0)}

Next, for each row corresponding to edge $\langle v,w\rangle$, compute the number of edges of the expanded $E$
incoming into $v$.

\X{C5}{=COUNTIF(C3,RC1)}

The final column of formulas computes the levels of vertices. If the
following formula is located in a row corresponding to an edge $\langle v,w\rangle$,
it computes the level of $v$.

\Y{C6}{=IF(RC1=R1C3,0,IF(ISERROR(RC3),1E300,\\1+MIN(OFFSET(R1C6,RC4-1,0,RC5))))}

First, the start vertex $s$ from \texttt{R1C3} is assigned level
0. Next, if $v$ is not $s$ and has no incoming edges, its level
becomes \texttt{1E300}. If $v$ has incoming edges, \texttt{OFFSET}
function creates a range \textit{in the present column}, encompassing
exactly all rows whose edges in columns \texttt{C1:C2} have the form
$\langle w,v\rangle.$ The formula then calculates \texttt{MIN} over
this range and adds 1. This is the level of $v$ and its value is then
recorded in column\texttt{C6}. Since then it becomes available for
computing levels of further vertices, allowing recursion. Note, that
the acyclicity of $G$ implies that we do not get cyclic references and
the computation is well-founded.

One can observe that the pattern of the references within column
\texttt{C5} reproduces exactly the expanded set of edges $E$ whose
traversal we perform.

What remains to be done is to sort column \texttt{C1} according to the
value in \texttt{C5}, replace all vertices of level
$\geq\mathtt{1E300}$ by \texttt{\#N/A!} and finally eliminate the
duplicates. Again, this is doable using relational operators.

An interesting observation is that such a spreadsheet can also perform
the test if the input graph is acyclic, but in a quite specific way. A
successful computation of the above spreadsheet determines that the
graph is indeed acyclic. If it is not, the spreadsheet responds with a
message that circular references are created.

\begin{figure*}[t]
\begin{verbatim}
cur_node:=s;
first_time:=true;
cur_parent:=NULL;
next_sibling:=NULL;
first_son:=the first son of cur_node;
visited:=<cur_node,first_time,cur_parent>;

repeat 2*|E| times
if first_time and first_son<>NULL 
  then {move to the first son}
    cur_parent:=cur_node;
    cur_node:=first_son;
    first_son:=the first son of cur_node;
    first_time:=[cur_node does not appear in visited];
    visited:=append(visited,<cur_node,first_time,cur_parent>);
elseif next_sibling<>NULL and ((first_time and first_son=NULL) or not(first_time)) 
  then {move to the next sibling}
    cur_node:=next_sibling;
    first_son:=the first son of cur_node;
    first_time:=[cur_node does not appear in visited];
    visited:=append(visited,<cur_node,first_time,cur_parent>);
else {backtrack}
    cur_node:=cur_parent;
    cur_parent:=the parent of cur_node when cur_node was new, from visited;
    first_son:=the first son of cur_node;
    first_time:=[cur_node does not appear in visited];
    visited:=append(visited,<cur_node,first_time,cur_parent>);
\end{verbatim}
\caption{The pseudo-code of the algorithm for DFS.}
\label{fig:dfs}
\end{figure*}

\subsection{Depth-first-search}

In an analogous situation, our present task is to determine the
ordering of the vertices of $G$ in which they are discovered by DFS
algorithm run on $G$ starting from $s$. We assume that the edges
of $E$ are listed in some order. This order determines the relative
order of the vertices reachable from any given vertex $v$.
\begin{figure*}[htb]
\centering
\includegraphics[scale=0.7]{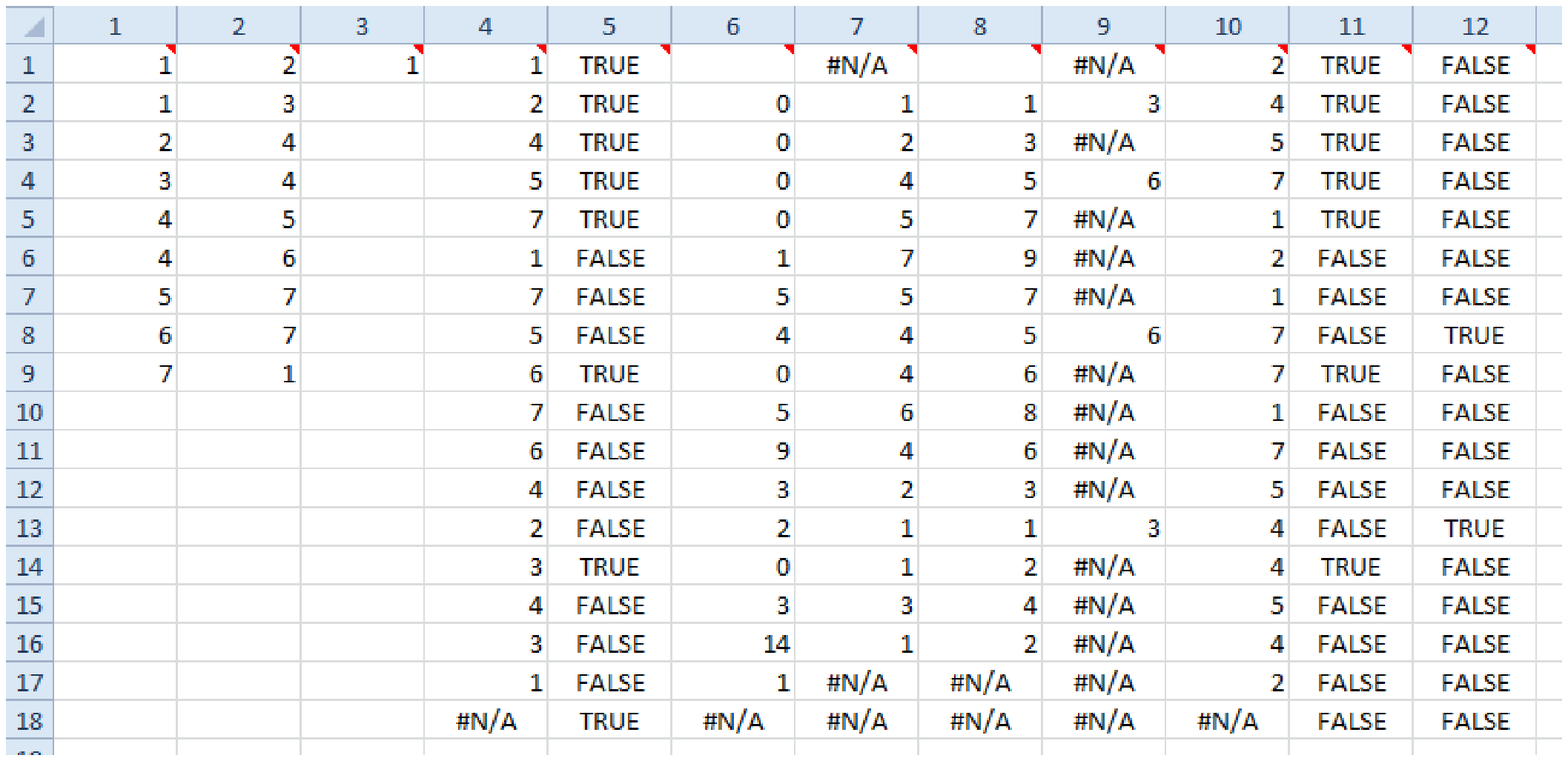}
\caption{Spreadsheet implementation of the Depth-First-Search graph
  traversal.}
\label{fig:DFS}
\end{figure*}

Now, we describe the iterative algorithm which performs DFS. We will
mimic its operation in a spreadsheet presented later, to determine the
order in which it visits the vertices.

At all times the algorithm keeps the following information: 
\begin{itemize}
\item the current vertex $\mathit{cur\_node}$,
\item Boolean $\mathit{first\_time}$ which gives the information if
  the present visit to $\mathit{cur\_node}$ is the first one or not,
\item the parent vertex $\mathit{cur\_parent}$ from which we have
  reached $\mathit{cur\_node}$,
\item the $\mathit{first\_son}$ of $\mathit{cur\_node}$,
\item the present $\mathit{next\_sibling}$ of $\mathit{cur\_node}$,
  which is the next vertex after $\mathit{cur\_node}$ among those
  reachable from $\mathit{cur\_parent}$,
\item the list $\mathit{visited}$ of all tuples
  $(\mathit{cur\_node},\mathit{first\_time},\mathit{cur\_parent})$
  encountered so far, in the order of appearance.
\end{itemize}

Each of the above vertices can be NULL.  In Fig. \ref{fig:dfs} we
present the pseudo-code of the algorithm for DFS.  The body of its
\texttt{repeat} statement is composed of a three-way conditional
statement.  Note that in all the three branches the last three
statements are exactly the same.  It is intentional, since this is not
a classic ``if'' statement but a construction composed of three
spreadsheet cells that implement the branches.  Even if the
factorisation of those three statements out of the conditional was
feasible in traditional programming languages, it would
\emph{complicate} not simplify the resulting spreadsheet.

Since the number of iterations of the main loop is fixed, it is
important to note what happens if the traversal is finished in fewer
than $2*|E|$ steps: the algorithm then attempts to backtrack from the
start vertex $s$ and makes $\mathit{cur\_node}$ \texttt{NULL}. From
that moment on, whatever is computed from \texttt{NULL} results in
\texttt{NULL}, so the rest of the output consists of \texttt{NULL}s.

This algorithm can be now translated into a spreadsheet. We do not
need expanded $E$ for that purpose, since we do not touch unreachable
vertices. The first step is to order the edges of $E$ by their first
coordinate. This does not alter the relative order of children of the
vertices.  We omit the formulas which do the sorting (see Section
\ref{sec:sorting}).

The columns of the spreadsheet hold the values of variables of the
above algorithm, each row corresponding to one iteration, and there
are a few columns with auxiliary computations.  Values of
$\mathit{cur\_node}$ are stored in column \texttt{C4},
$\mathit{first\_time}$ in \texttt{C5}, $\mathit{cur\_parent}$ in
\texttt{C6}, $\mathit{next\_sibling}$ in \texttt{C8},
$\mathit{first\_son}$ in \texttt{C9}. Columns \texttt{C11} and
\texttt{C12} calculate the logical values of tests in the \texttt{if
  \ldots then} and \texttt{elseif \ldots then} statements of the
algorithm, computed according to the values at the end of the
iteration. Finally, column \texttt{C13} holds numbers of consecutive
rows, which helps in calculations.

First, $\mathit{cur\_node}$ is determined in column \texttt{C4} depending on the logical
tests computed at the end of the previous iteration.

\Y{C4}{=IF(R[-1]C11,R[-1]C10,\\IF(R[-1]C12,R[-1]C9,R[-1]C7))}

The special formula in row 1 of that column is the initialization of
$\mathit{cur\_node}$.

\Y{R1C4}{=R1C3}

The next column consists of formulas checking if the present value of
$\mathit{cur\_node}$ appears in the history of that variable. \texttt{MATCH} returns an
error when it is not.

\X{C5}{=ISERROR(MATCH(RC4,R1C4:R[-1]C4,0))}

The next column of formulas computes an auxiliary value: the row
number of the first time the present value of $\mathit{cur\_node}$ was
encountered. This first time is always distinguished by \texttt{TRUE}
accompanying it in column \texttt{C5}. Because the present value of
$\mathit{cur\_node}$ was a new one at most once, the sum of row numbers in column
\texttt{C13} yields the step number when it
was new, or 0 if it was never new in the past (because it is new now).

\X{C6}{=SUMIFS(R1C13:R[-1]C13,R1C4:R[-1]C4,RC4,\\R1C5:R[-1]C5,TRUE)}

We initialize this column by leaving the first cell empty.

\X{R1C6}{}

Now we compute the value of $\mathit{cur\_parent}$ for the present
$\mathit{cur\_node}$. In the case we are doing backtrack, the parent
of $\mathit{cur\_node}$ is found in the history, by looking up the
moment when $\mathit{cur\_node}$ was new. This position has been
computed in \texttt{RC6}, and we take the corresponding value from the
history of $\mathit{cur\_parent}$ in column \texttt{C7}.

\X{C7}{=IF(R[-1]C11,R[-1]C4,IF(R[-1]C12,\\R[-1]C7,INDEX(R1C7:R[-1]C7,RC6)))}

For the start vertex in the first row we initialize $\mathit{cur\_parent}$ as
\texttt{NULL}.  

\X{R1C7}{=NA()} 

Next we compute another auxiliary value: the number of the row of the
input relation $E$ in which the edge $\langle
\mathit{cur\_parent},\mathit{cur\_node}\rangle$ appears. It appears
there exactly once, so again we sum the row numbers from column
\texttt{C13}.

\X{C8}{=SUMIFS(C13,C1,RC7,C2,RC4)} 

However, at the beginning the pair $\langle
\mathit{cur\_parent},\mathit{cur\_node}\rangle$ is not an edge, so we
initialize leaving the first cell in this column empty.

\X{R1C8}{}

Referring to the auxiliary value computed in \texttt{RC8}, the
following formula checks if in the edge of $E$ following $\langle
\mathit{cur\_parent},\mathit{cur\_node}\rangle$ in the listing, the
vertex $\mathit{cur\_parent}$ is still the parent. If it is so, the
second coordinate of that edge is the present next sibling of
$\mathit{cur\_node}$. Otherwise we set the next sibling to
\texttt{NULL}.

\X{C9}{=IF(INDEX(C1,RC8+1)=RC7,INDEX(C2,RC8+1),\\NA())} 

Of course, the start vertex has no next sibling, so we initialize this
column with \texttt{NULL} in the first row.

\X{R1C9}{=NA()}

In column \texttt{C10} we find the first edge of $E$ whose first
coordinate is the present $\mathit{cur\_node}$, and return the second
coordinate as the first son $\mathit{first\_son}$ of
$\mathit{cur\_node}$. If there is no such edge in $E$,
$\mathit{first\_son}$ becomes \texttt{NULL}. This column does not
require any specific initialization in the first row.

\X{C10}{=IFERROR(INDEX(C2,MATCH(RC4,C1,0)),NA())}

The last three columns compute the test results for \texttt{if \ldots
  then} and \texttt{elseif \ldots then}, to be used in the next
iteration, and the row number.

\X{C11}{=AND(RC5,NOT(ISNA(RC[-2])))}

\X{C12}{=AND(NOT(ISNA(RC8)),OR(NOT(RC5),\\AND(RC5,ISNA(RC9))))}

\X{C13}{=ROW()}

Now it is sufficient to remove duplicates from column \texttt{C4} to
get the order in which the initial algorithm discovers new vertices of
the input graph. We omit the formulas for doing that.

\subsubsection{Example application}
The discussed graph traversal algorithms can be used to implement
recursive queries over hierarchies.  Hierarchical data is ubiquitous.  In shops
we have product categorization with some categories being subcategories
of other.  Companies have numerous corporate hierarchies of employees
(manager-subordinate), organizational units (unit-subunit) or projects
(project-subproject) etc. Although not covered in relational algebra,
recursive queries to data representing hierarchies are often encountered in practice.
In order to query such relations recursive
queries have been introduced.  First, they emerged in Oracle database
in the form of the famous \texttt{CONNECT BY} clause, and then SQL:1999
adopted them as recursive common table expressions.

A spreadsheet with an example implementation of the described graph
algorithms is available online as a supplementary material (see Section~\ref{webPage}).
The spreadsheet also includes an example of how the described techniques
can be used to implement a typical query on the sample Oracle
manager-subordinate hierarchy:

\begin{verbatim}
SELECT ename, level
FROM emp
START WITH empno = 7839
CONNECT BY PRIOR empno = mgr;
\end{verbatim}

\section{Conclusions, further research}

We have demonstrated that SQL can be automatically translated into
spreadsheet code, including NULL values.  Thus, we have shown the
power of the spreadsheet paradigm, which subsumes the paradigm of
relational databases.

Apart from SQL, we have also implemented a few specific
algorithms: a linearithmic sorting procedure and two graph traversing
algorithms: BFS and DFS.

As the next steps:
\begin{itemize}
\item We plan to develop optimizations for SQL queries translated into
  spreadsheet.
\item We plan to investigate whether spreadsheets can naturally
  implement other models of databases, like semi-structured or
  object-relational ones.
\item Google spreadsheets provide the \texttt{QUERY} function,
  capable of expressing a limited fragment of SQL, as well as
  \texttt{SORT}, \texttt{FILTER} and \texttt{UNIQUE} functions, of
  similar roles. If a Google spreadsheet with those functions is
  downloaded as \texttt{xlsx} or \texttt{ods} file, these functions
  are not recognized by other spreadsheet systems and produce
  errors. We plan to use our implementations of algebra operators and
  experience with SQL translator to construct a translator capable of
  producing fully functional spreadsheet files from those downloaded from
  Google docs.
\end{itemize}

\section{Software availability}\label{webPage}

The Web page of the present paper is
\url{hhtp://www.mimuw.edu.pl/~jty/Translating/}, from which the
software described in this paper can be accessed, including:
\begin{itemize}
\item the SQL to \textit{Excel} translator, described in Section
  \ref{SQL2XLSX}. This tool is independently accessible from
  Sourceforge
  \url{http://sourceforge.net/projects/sqltoalgebra/?source=directory},
\item  \textit{Excel} implementation of a sorting algorithm, working in
  time $O(n\log^2 n)$, described in Section \ref{Sort}, and
\item \textit{Excel} implementations of BFS and DFS. 
\end{itemize}

\section*{Acknowledgments} 
The present paper is a significantly rewritten version of the
conference paper \cite{Tyszkiewicz}. It has benefited substantially
from the suggestions of several anonymous conference reviewers. We
would like to thank Jan Van den Bussche for many valuable discussions
and advices.

The research of J.S. and J.Ty. has been funded by Polish National
Science Centre (Narodowe Centrum Nauki).

\bibliographystyle{IEEEtran} 
\bibliography{IEEEabrv,SIGMOD}

\begin{thebibliography}{10}
\providecommand{\url}[1]{#1}
\csname url@samestyle\endcsname
\providecommand{\newblock}{\relax}
\providecommand{\bibinfo}[2]{#2}
\providecommand{\BIBentrySTDinterwordspacing}{\spaceskip=0pt\relax}
\providecommand{\BIBentryALTinterwordstretchfactor}{4}
\providecommand{\BIBentryALTinterwordspacing}{\spaceskip=\fontdimen2\font plus
\BIBentryALTinterwordstretchfactor\fontdimen3\font minus
  \fontdimen4\font\relax}
\providecommand{\BIBforeignlanguage}[2]{{%
\expandafter\ifx\csname l@#1\endcsname\relax
\typeout{** WARNING: IEEEtran.bst: No hyphenation pattern has been}%
\typeout{** loaded for the language `#1'. Using the pattern for}%
\typeout{** the default language instead.}%
\else
\language=\csname l@#1\endcsname
\fi
#2}}
\providecommand{\BIBdecl}{\relax}
\BIBdecl

\bibitem{Gates99}
B.~Gates, ``Investing in research, {SIGMOD} {C}onference 1998 {K}eynote
  {S}peech, video,'' \emph{ACM SIGMOD Digital Symposium Collection}, vol.~1,
  no.~2, 1999.

\bibitem{MrExcel:Join}
\BIBentryALTinterwordspacing
SirMille. (2012, February) inner/outer/full join tables? [Online]. Available:
  \url{http://www.mrexcel.com/forum/excel-questions/612236-inner-outer-full-jo%
in-tables.html}
\BIBentrySTDinterwordspacing

\bibitem{MrExcel:social}
\BIBentryALTinterwordspacing
xil. (2012, May) Connecting list of users with list of companies - please help.
  [Online]. Available:
  \url{http://www.mrexcel.com/forum/excel-questions/704217-combining-two-lists%
-into-one-big-list.html}
\BIBentrySTDinterwordspacing

\bibitem{MrExcel:combining}
\BIBentryALTinterwordspacing
stefanaalten. (2013, May) Combining two lists into one big list. [Online].
  Available:
  \url{http://www.mrexcel.com/forum/excel-questions/704217-combining-two-lists%
-into-one-big-list.html}
\BIBentrySTDinterwordspacing

\bibitem{MrExcel:v-lookup}
\BIBentryALTinterwordspacing
mzalikhan. (2012, Apr.) {Double V-look up}. [Online]. Available:
  \url{\url{http://www.mrexcel.com/forum/excel-questions/625750-double-v-look-%
up.html}}
\BIBentrySTDinterwordspacing

\bibitem{MrExcel:MsQuery}
\BIBentryALTinterwordspacing
potter\_ricky. (2014, March) Join two queries using msquery, returning into an
  excel workbook. [Online]. Available:
  \url{http://www.mrexcel.com/forum/excel-questions/765490-join-two-queries-us%
ing-msquery-returning-into-excel-workbook.html}
\BIBentrySTDinterwordspacing

\bibitem{Google:QL}
\BIBentryALTinterwordspacing
G.~{I}nc. (accessed 2014-05-21) Query {L}anguage {R}eference ({V}ersion 0.7).
  [Online]. Available:
  \url{https://developers.google.com/chart/interactive/docs/querylanguage}
\BIBentrySTDinterwordspacing

\bibitem{Tyszkiewicz}
\BIBentryALTinterwordspacing
J.~Tyszkiewicz, ``Spreadsheet as a relational database engine,'' in
  \emph{Proceedings of the 2010 ACM SIGMOD International Conference on
  Management of data}, ser. SIGMOD '10.\hskip 1em plus 0.5em minus 0.4em\relax
  New York, NY, USA: ACM, 2010, pp. 195--206. [Online]. Available:
  \url{http://doi.acm.org/10.1145/1807167.1807191}
\BIBentrySTDinterwordspacing

\bibitem{Excel}
{Microsoft Corporation}. (accessed 2014/05/21) {Excel Home Page - Microsoft
  Office Online}.
  \url{http://office.microsoft.com/en-us/excel-help/excel-functions-alphabetic%
al-list-HA010277524.aspx?CTT=1}.

\bibitem{icde}
B.~Liu and H.~V. Jagadish, ``A spreadsheet algebra for a direct data
  manipulation query interface,'' in \emph{ICDE '09: Proceedings of the 2009
  IEEE International Conference on Data Engineering}.\hskip 1em plus 0.5em
  minus 0.4em\relax Washington, DC, USA: IEEE Computer Society, 2009, pp.
  417--428.

\bibitem{2}
A.~Witkowski, S.~Bellamkonda, T.~Bozkaya, G.~Dorman, N.~Folkert, A.~Gupta,
  L.~Shen, and S.~Subramanian, ``Spreadsheets in {RDBMS} for {OLAP},'' in
  \emph{SIGMOD '03: Proceedings of the 2003 ACM SIGMOD international conference
  on Management of data}.\hskip 1em plus 0.5em minus 0.4em\relax New York, NY,
  USA: ACM, 2003, pp. 52--63.

\bibitem{3}
A.~Witkowski, S.~Bellamkonda, T.~Bozkaya, N.~Folkert, A.~Gupta, L.~Sheng, and
  S.~Subramanian, ``Business modeling using {SQL} spreadsheets,'' in \emph{VLDB
  '2003: Proceedings of the 29th international conference on Very large data
  bases}.\hskip 1em plus 0.5em minus 0.4em\relax VLDB Endowment, 2003, pp.
  1117--1120.

\bibitem{4}
A.~Witkowski, S.~Bellamkonda, T.~Bozkaya, A.~Naimat, L.~Sheng, S.~Subramanian,
  and A.~Waingold, ``Query by {Excel},'' in \emph{VLDB '05: Proceedings of the
  31st international conference on Very large data bases}.\hskip 1em plus 0.5em
  minus 0.4em\relax VLDB Endowment, 2005, pp. 1204--1215.

\bibitem{6}
L.~V.~S. Lakshmanan, S.~N. Subramanian, N.~Goyal, and R.~Krishnamurthy, ``On
  query spreadsheets,'' in \emph{ICDE}.\hskip 1em plus 0.5em minus 0.4em\relax
  IEEE Computer Society, 1998, pp. 134--141.

\bibitem{1140346}
R.~Abraham and M.~Erwig, ``Type inference for spreadsheets,'' in \emph{PPDP
  '06: Proceedings of the 8th ACM SIGPLAN Symposium on Principles and Practice
  of Declarative Programming}.\hskip 1em plus 0.5em minus 0.4em\relax New York,
  NY, USA: ACM, 2006, pp. 73--84.

\bibitem{Forms}
M.~M. Burnett, J.~W. Atwood, R.~W. Djang, J.~Reichwein, H.~J. Gottfried, and
  S.~Yang, ``Forms/3: A first-order visual language to explore the boundaries
  of the spreadsheet paradigm,'' \emph{J. Funct. Program.}, vol.~11, no.~2, pp.
  155--206, 2001.

\bibitem{944721}
S.~P. Jones, A.~Blackwell, and M.~Burnett, ``A user-centred approach to
  functions in {E}xcel,'' in \emph{ICFP '03: Proceedings of the Eighth ACM
  SIGPLAN International Conference on Functional Programming}.\hskip 1em plus
  0.5em minus 0.4em\relax New York, NY, USA: ACM, 2003, pp. 165--176.

\bibitem{1}
M.~Kassoff, L.-M. Zen, A.~Garg, and M.~Genesereth, ``{PrediCalc}: a logical
  spreadsheet management system,'' in \emph{VLDB '05: Proceedings of the 31st
  international conference on Very large data bases}.\hskip 1em plus 0.5em
  minus 0.4em\relax VLDB Endowment, 2005, pp. 1247--1250.

\bibitem{885148}
R.~Mittermeir and M.~Clermont, ``Finding high-level structures in spreadsheet
  programs,'' in \emph{WCRE '02: Proceedings of the Ninth Working Conference on
  Reverse Engineering (WCRE'02)}.\hskip 1em plus 0.5em minus 0.4em\relax
  Washington, DC, USA: IEEE Computer Society, 2002, p. 221.

\bibitem{63244}
B.~Ronen, M.~A. Palley, and J.~Henry C.~Lucas, ``Spreadsheet analysis and
  design,'' \emph{Commun. ACM}, vol.~32, no.~1, pp. 84--93, 1989.

\bibitem{1194878}
D.~Wakeling, ``Spreadsheet functional programming,'' \emph{J. Funct. Program.},
  vol.~17, no.~1, pp. 131--143, 2007.

\bibitem{184736}
A.~G. Yoder and D.~L. Cohn, ``Architectural issues in spreadsheet languages,''
  in \emph{Proceedings of the international conference on Programming languages
  and system architectures}.\hskip 1em plus 0.5em minus 0.4em\relax New York,
  NY, USA: Springer-Verlag New York, Inc., 1994, pp. 245--258.

\bibitem{DBLP:conf/iccl/YoderC94}
------, ``Real spreadsheets for real programmers,'' in \emph{ICCL}, H.~E. Bal,
  Ed.\hskip 1em plus 0.5em minus 0.4em\relax IEEE Computer Society, 1994, pp.
  20--30.

\bibitem{DB-implementation}
H.~Garcia-Molina, J.~D. Ullman, and J.~Widom, \emph{Database System
  Implementation}.\hskip 1em plus 0.5em minus 0.4em\relax Prentice-Hall, 2000.

\bibitem{VandenBussche}
J.~V. den Bussche and S.~Vansummeren, ``Translating {SQL} into the relational
  algebra,'' Lecture material, Universiteit Limburg, lecture INFO-H-417:
  Database Systems Architecture.

\bibitem{DBLP:conf/icde/GrefenB94}
P.~W. P.~J. Grefen and R.~A. de~By, ``A multi-set extended relational algebra -
  a formal approach to a practical issue,'' in \emph{ICDE}.\hskip 1em plus
  0.5em minus 0.4em\relax IEEE Computer Society, 1994, pp. 80--88.

\end{thebibliography}

\end{document}